%% file: main_arxiv.tex
\documentclass[reprint, superscriptaddress, amsmath,amssymb,aps,]{revtex4-2}
\usepackage{amsmath,amsthm,amssymb}
\usepackage{wasysym}
\usepackage{float}
\usepackage{graphicx}
\usepackage{dcolumn}
\usepackage{bm}
\usepackage{url}
\usepackage{hyperref}
\usepackage{placeins}
\usepackage{color}
\usepackage{cleveref} 
\crefformat{footnote}{#2\footnotemark[#1]#3}

\newcommand\blue[1]{\textcolor{blue}{#1}}

\begin{document}

\preprint{APS/123-QED}

\title{Additive Manufacturing of functionalised atomic vapour cells for next-generation quantum technologies}

\author{F. Wang}
\affiliation{Faculty of Engineering, University of Nottingham,  University Park, Nottingham, NG7 2RD, UK}
\email{ppzfw@exmail.nottingham.ac.uk}
%\email{nathan.cooper@nottingham.ac.uk}
\author{N. Cooper}
\affiliation{School of Physics and Astronomy, University of Nottingham, University Park, Nottingham, NG7 2RD, UK}
\email{nathan.cooper@nottingham.ac.uk}
\author{Y. He$^1$}
\affiliation{Faculty of Engineering, University of Nottingham,  University Park, Nottingham, NG7 2RD, UK}
\author{B. Hopton}
\affiliation{School of Physics and Astronomy, University of Nottingham, University Park, Nottingham, NG7 2RD, UK}
\author{D. Johnson}
\affiliation{School of Physics and Astronomy, University of Nottingham, University Park, Nottingham, NG7 2RD, UK}
\author{P. Zhao}
\affiliation{Faculty of Engineering, University of Nottingham,  University Park, Nottingham, NG7 2RD, UK}
\author{T. M. Fromhold}
\affiliation{School of Physics and Astronomy, University of Nottingham, University Park, Nottingham, NG7 2RD, UK}
\author{C. J. Tuck}
\affiliation{Faculty of Engineering, University of Nottingham,  University Park, Nottingham, NG7 2RD, UK}
\author{R. Hague}
\affiliation{Faculty of Engineering, University of Nottingham,  University Park, Nottingham, NG7 2RD, UK}
\author{R. D. Wildman}
\affiliation{Faculty of Engineering, University of Nottingham,  University Park, Nottingham, NG7 2RD, UK}
\author{L. Turyanska}
\affiliation{Faculty of Engineering, University of Nottingham,  University Park, Nottingham, NG7 2RD, UK}
\author{L. Hackerm\"{u}ller}
\affiliation{School of Physics and Astronomy, University of Nottingham, University Park, Nottingham, NG7 2RD, UK}

\date{\today}

\begin{abstract}
Atomic vapour cells are an indispensable tool for quantum technologies (QT), but potential improvements are limited by the capacities of conventional manufacturing methods. Using an additive manufacturing (AM) technique - vat polymerisation by digital light processing -  we demonstrate, for the first time,  a 3D-printed glass vapour cell. The exploitation of AM capacities allows intricate internal architectures, overprinting of 2D optoelectronical materials to create integrated sensors and surface functionalisation, while also showing the ability to tailor the optical properties of the AM glass by in-situ growth of gold nanoparticles. The produced cells achieve ultra-high vacuum of $2 \times 10^{-9}$\,mbar and enable Doppler-free spectroscopy; we demonstrate laser frequency stabilisation as a QT application. These results highlight the transformative role that AM can play for QT in enabling compact, optimised and integrated multi-material components and devices.
\end{abstract}

\maketitle

\section{Introduction}

Growing understanding of fundamental quantum processes offers benefits in multiple sectors, from medicine\,\cite{aslam2023quantum,SPMIC} and sensing applications\,\cite{arita2013laser,Bongs2019,Lee2023}, positioning and security\,\cite{feng2019review, Stray2022} to quantum computing\,\cite{Evered2023}. However, realising these benefits requires innovations in quantum technology (QT) hardware, which must become smaller, lighter, cheaper, and better suited to the needs of specific applications. Additive manufacturing (AM) has the potential to transform QT\,\cite{Ruchka2022, vovrosh2018} by enabling miniaturised, robust devices that can be fabricated on demand with advanced architectures for enhanced operation, such as the innovative ultra-high vacuum apparatus\,\cite{cooper2021printedchamber} previously reported. AM also presents opportunities for integration, both of novel hardware with existing equipment\,\cite{madkhaly2021printedoptics,Staerkind2023} or of functionalised components within a larger device.

A critical QT component is the atomic vapour cell\,\cite{Liew2004,Maurice2022}, it underpins many technologies, spanning laser frequency stabilisation\,\cite{Cooper2023, Martinez2023}, atom, molecule and ion-trapping\,\cite{Vilas_2022,HAFFNER_2008,Marciniak2023} thermal vapour magnetometers\,\cite{Fabricant_2023}, medical imaging\,\cite{aslam2023quantum,SPMIC} and industrial scanning applications\,\cite{pross2005second,behzadirad2021advanced}. However, vapour cells are difficult to manufacture, rely on the art of glass-blowing and are consequently limited in size and shape. They are usually cylindrical with dimensions on the order of centimetres, and have only limited scope for miniaturisation or customisation. 

Planar lithographic processes\,\cite{petremand2012microfabricated,shah2007subpicotesla} and  microelectromechanical systems (MEMS) with anodic bonding\,\cite{Gong2006,serf,Lee2004} can produce microfabricated cells with great potential for miniaturisation, but lack 3D-versatility and offer only two optical interfaces, which is insufficient for some applications; nor do they have the same transformative potential for integration and customisation as AM. 

Additive manufacturing can provide the means to overcome these limitations. AM of glass was demonstrated using fused filament fabrication of molten glass\,\cite{luo2017additive,inamura2018additive}, selective laser melting with silica powder\,\cite{datsiou2019additive,lei2019additive} and direct ink writing using silica sol-gel inks\,\cite{sasan2020additive}. However, significant technological challenges such as limited building resolution, evident surface roughness, cracking and low optical transparency have also been identified. Glass printing via the use of photocurable polymer / silica nanoparticle composites has led to optically useful constructs\,\cite{kotz2017three,cooperstein2018additive,moore2020three,toombs2022volumetric,kotz2021two,han2019rapid}, but to date, the use of these formulations has been limited to structural ornaments, classical optics and fluidics. 

\begin{figure*}[ht]
\centering
\includegraphics[width=0.8 \linewidth]{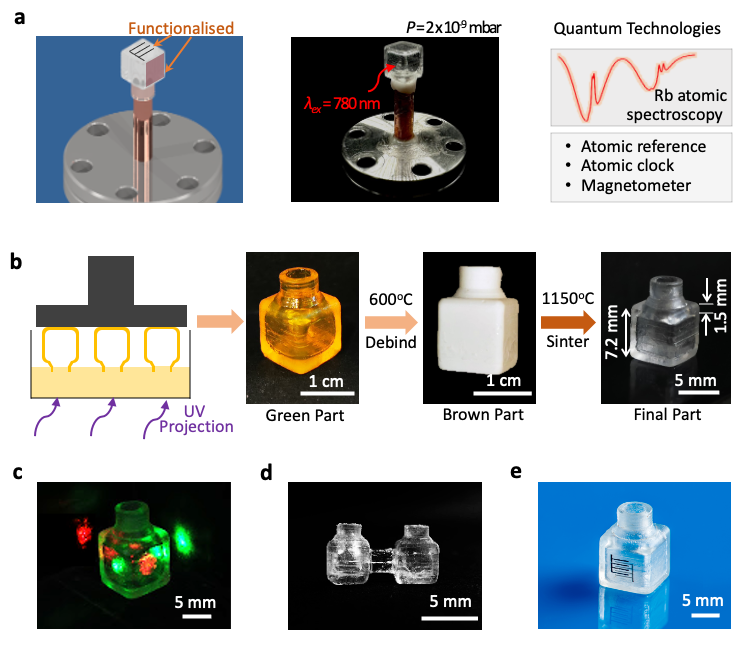}
\caption{\textbf{Additive manufacturing of a vapour cell for QT applications.} \textbf{a} Schematic diagram of the AM vapour cell for QT applications. Left: Illustration of the design of the cell mounted on an ultra-high vacuum flange via a copper tube. Middle: The photo of  AM cell mounted for pumping to a pressure of $2\times 10^{-9}$ mbar and loading with Rb atomic vapour. Right: Doppler-free pump-probe absorption spectroscopy of Rb vapour in the AM vapour cell and applications for QT. \textbf{b} Vat polymerisation of glass vapour cells via DLP and optical images of the printed green, brown and final parts following debinding and sintering steps. \textbf{c} Photograph of the printed cell with red and green laser beams propagating through the cell. \textbf{d} New structures are achievable: photograph of two inter-connected cells. \textbf{e} Integrated functionalisation: Inkjet printed graphene and silver inter-digit electrodes for direct photon detection on the cell.}
\label{fig:1}
\end{figure*}

Here we provide the first demonstration of an ultra-high vacuum (UHV) compatible AM glass QT component and show that, by modifying the inks and printing procedures, transparent printed structures with integrated active elements (i.e. electronics, photodetectors) can be realised. These structures are of high optical quality, have the potential to transform vapour cell based QTs and open a route to the scalable manufacture of key components for spectroscopy and quantum sensing (Figure \ref{fig:1}a)).

%%%%% Fig. 1

%%%%% Fig. 2

\begin{figure*}[ht]
\centering
\includegraphics[width=0.8 \linewidth]{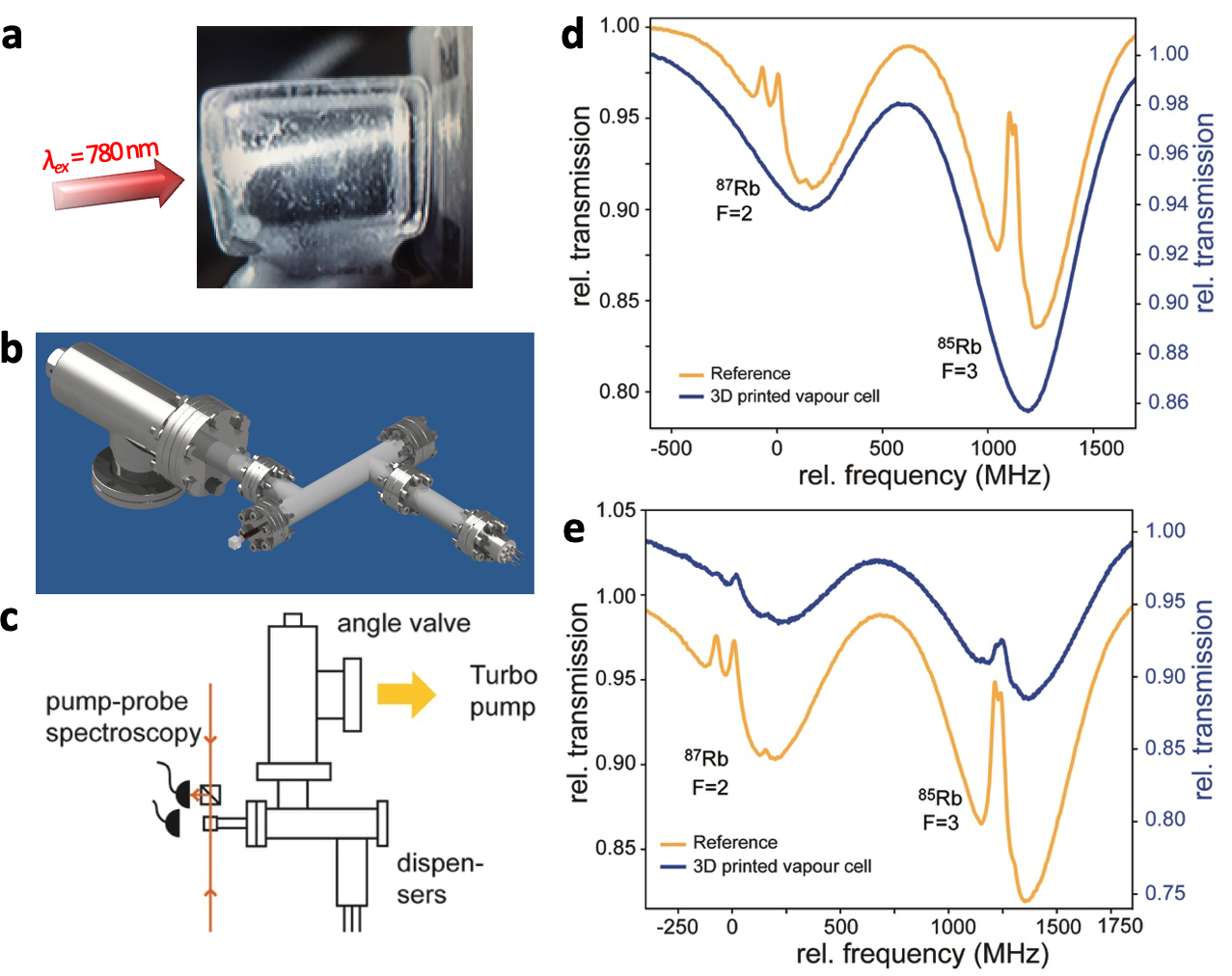}
\caption{\textbf{Absorption spectroscopy of Rb vapour in the AM vapour cell.} \textbf{a} Rb fluorescence image inside the cell under exposure to $\lambda_{ex} = 780$\,nm laser light. \textbf{b} and \textbf{c} CAD model and a scheme of the vacuum and optical setups. \textbf{d} Single beam absorption spectroscopy. The scale of the reference signal (yellow) is vertically offset by 0.02 for clarity. The frequency scale in d and e is set relative to the $^{87}$Rb $F = 2 \rightarrow F' = 2 \times 3$ transition. \textbf{e} Doppler-free pump-probe absorption spectroscopy. For clarity, the scale of the reference signal (yellow) is vertically offset by -0.05.}
\label{fig:spectroscopy}
\end{figure*}

We describe the fabrication of glass vapour cells with sizes $< 1~$cm$^3$ by a vat polymerisation method based on digital light processing (DLP) of high loading silica nanoparticles resins\,(Figure \ref{fig:1}b). As a QT application, the cells are used to demonstrate atomic spectroscopy and show excellent vacuum performance, with pumped cells reaching pressures below $2\times 10^{-9}$\,mbar and sustaining this pressure throughout the experiments. The cells were thermally tested up to $150^{\circ}$C and were not noticeably degraded by exposure to high-temperature Rb vapour. 

We performed atomic vapour spectroscopy within the cells and demonstrated its application to laser frequency stabilisation\,\cite{CooperDualFreq} on the D2 line of $^{85}$Rb. 

The fabrication methodology is further applied to create complex cell geometries, while in-situ growth of metal nanoparticles is used to selectively tune the optical transmission of the produced glass, providing a clear demonstration of the versatility and transformative potential of such a method for creating optical materials for QT. 

These functionalised cells enable exciting applications, e.g. for the creation of sub-cm sized optically pumped magnetometers incorporating magnetic-field shielding conductors and photodetectors directly printed on the cell, thereby providing a significant jump in system flexibility and imaging resolution for non-invasive magnetoencephalography\,\cite{Brookes2022}.

\section{Results and Discussion}
\subsection{Additive manufacturing of a vapour cell}

To manufacture the glass vapour cell (Figure \ref{fig:1}), a resin was formulated based on\,\cite{kotz2017three} containing fumed silica nanoparticles with an average size of 40 nm, dispersed in a mixture of 2-hydroxyethyl methacrylate (HEMA), tetra(ethylene glycol) diacrylate (TEGDA) and phenoxyethanol (POE). HEMA forms a solvation layer on the nanoparticles, hence allowing high silica loading of the resin. TEGDA is used to improve cross-linking, thus strengthening the mechanical structure of the printed part. A concentration of 50 wt\% silica nanoparticles was achieved following a multi-step homogenising process \cite{Cai2020}. The viscosity of the resin was measured to be 297\,mPa$\cdot$s at a shear rate of 1000\,s$^{-1}$. For demonstration purposes, and to facilitate comparison with existing vapour cells used in commercial QTs, our cells have a cuboid shape featuring two pairs of parallel optical interfaces, in contrast to the single pair of parallel interfaces in wafer-based cells\,\cite{Mitchell_2022} and most glass-blown cells.
The intended cell dimensions were  7.2\,mm (L) $\times ~7.2$\,mm (W) $\times ~7.2$\,mm (H) with 1.5\,mm wall thickness; the initial print was therefore 1\,cm$^3$ with 2 mm thick walls, to allow for shrinkage. An inlet on one side (Figure \ref{fig:1}b) enables vapour loading and connection to our experimental apparatus. 

To optimise the degree of curing during printing and obtain a high shape fidelity between the printed part and the design, the exposure time of each layer and the concentration of the absorber was adjusted by measuring the curing depth with different absorber concentrations (Appendix A1, Figure \ref{fig:S1}). The optimal degree of polymerisation, curing depth and shape fidelity, was achieved with 6.5\,s exposure time, 45\,mW/cm$^2$ intensity and 0.035\,wt\% absorber content. A simulation of light-scattering from the silica nanoparticles showed an anisotropic distribution, however an isotropic degree of polymerisation can be achieved using this resin composition (Appendix A2, Figure \ref{fig:S2}). \\
The residual uncured resin was removed in washing and post-curing steps, resulting in a green part with good mechanical strength. Thermal debinding followed, with a gradual and uniform release of the internal mechanical stress, which was tailored to avoid cracking or collapse of the structure. The resulting debound porous brown part maintained the same geometry as the green part without volume shrinkage, but with a 51\% weight loss as a consequence of the debinding process.\\
To eliminate porosity, the part was sintered in an inert argon atmosphere at a temperature of $T = 1150\,^\circ$C for 12 h. This process allows the silica nanoparticles to merge and form amorphous glass (Figure \ref{fig:S3}). We note that the argon atmosphere prevents the risk of crystallisation, which was previously observed in synthesis of parts in the presence of oxygen\,\cite{Yong2002}. X-ray diffraction (XRD) analysis confirmed the amorphous structure of the printed material (Figure \ref{fig:S3}a)\,\cite{Milonjic2007}. The density of the sintered part is 2.2\,gcm$^{-3}$ and the sintered part had no observable porosity (see SEM images in Figure \ref{fig:S3}b)). The cells printed using this formulation show high mechanical stability under external pressures and good optical flatness/transparency (see below) and enabled us to achieve and retain vacuum down to $2 \times 10^{-9}$\,mbar. 

For our formulation, we observed $\sim 27\%$ linear shrinkage, both horizontally and vertically. The shrinkage of the part matched the estimation based on the solid weight concentration of the silica\,\cite{Cai2020} and was accounted for in the design. The two pairs of parallel transparent flat surfaces of the cells allow two perpendicular beams to pass through with minimal distortion (Figure \ref{fig:S1}c). An additional polishing step (Grit 1000 polishing paper) was introduced to reduce outer wall roughness.

\subsection{QT sensing with an additively manufactured cell}

To demonstrate the suitability of the printed cells for quantum technology applications, we performed rubidium spectroscopy in the cell. For this, the cell was mounted to an ultra-high vacuum (UHV) flange via an annealed copper tube (Figure \ref{fig:spectroscopy}a) and connected to a vacuum test system (Figure \ref{fig:spectroscopy}b). The cells were leak tested and found to retain pressures down to $2\times 10^{-9}$\,mbar. The annealed copper tube allows for a separation of the cells from the vacuum apparatus after evacuation and use of the cell as a stand-alone device, via a standard copper pinch-off technique. The cell was then filled with a mixture of $^{85}$Rb and $^{87}$Rb via atomic dispensers. The geometry of the cell is also compatible with alternative filling methods, reported previously\,\cite{Dispenserfilling,Mitchell_2022}.
The cell performance was first characterised through single-pass absorption spectroscopy. A single-mode, Gaussian laser beam with a diameter of 1.25 mm, a wavelength of 780 nm and a power of 0.06 mW was passed through the printed cell and detected on a photodiode (see Figure \ref{fig:spectroscopy}c). The transmission through the cell without Rb vapour is $>90\,$\%. 

The frequency of the laser light was scanned across the D2 resonance of rubidium 85 and fluorescence observed (inset of Figure \ref{fig:spectroscopy}a). The single-pass transmission spectrum (Figure \ref{fig:spectroscopy}d) displays the typical Voigt-profiles with all expected features. The yellow line shows Doppler-free absorption spectroscopy for comparison - this was obtained with a standard, commercial 75 mm long glass cell. The Doppler-valleys corresponding to the D2 absorption lines of $^{85}$Rb and $^{87}$Rb are clearly resolved, showing single pass absorption along the $^{87}$Rb $F = 2$ transition manifold and the $^{87}$Rb $F = 3$ transition manifold (Figure \ref{fig:spectroscopy}d). The frequency axis in Figure 3e is given relative to the $^{87}$Rb $F = 2 \rightarrow F’ = 2\times3$ transition. As an example of QT-sensing, Doppler-free saturated absorption spectroscopy was demonstrated with the cell at room temperature. Counter-propagating linearly polarised beams with powers of 0.06 mW (probe) and 0.3 mW (pump) were set up as shown in Figure \ref{fig:spectroscopy}c. The transmission of the probe beam (blue line in Figure \ref{fig:spectroscopy}e) shows the Doppler-free spectrum that allows the resolution of the hyperfine states, $^{87}$Rb $F = 2 \rightarrow F' = 1,2,3$ and $^{85}$Rb $F = 3 \rightarrow F' = 2,3,4$ (Figure \ref{fig:cellfabrication}e).

%%%%% Fig. 4

\begin{figure}[ht]
\centering
\includegraphics[width=\linewidth]{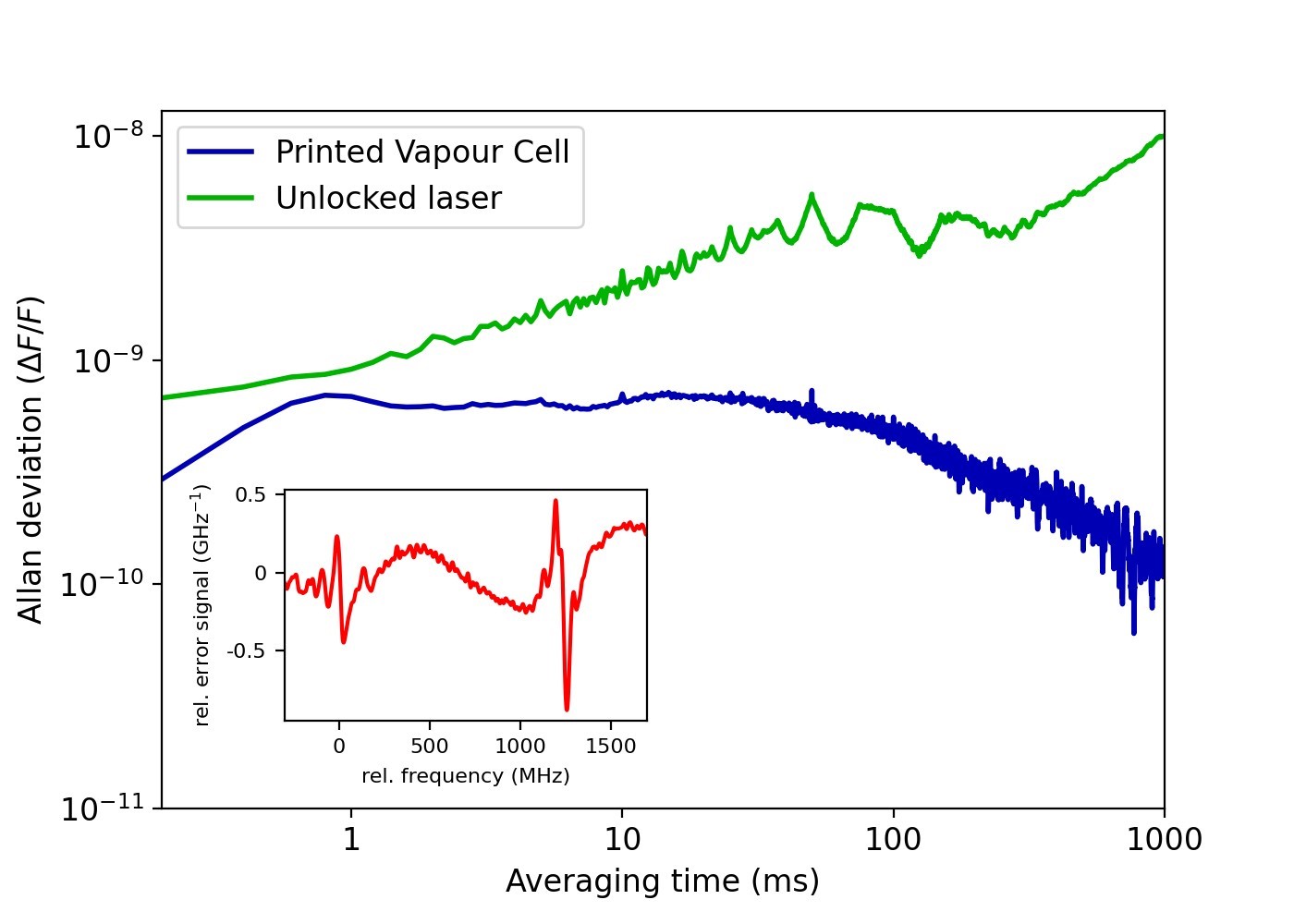}
\caption{\textbf{Allan deviation of the unlocked laser, and with the laser locked using the printed vapour cell.} Inset: The error signal (red) is shown for reference.}
\label{fig:Allan_deviation}
\end{figure}

%%%%% Fig. 4

\begin{figure*}[ht]
\centering
\includegraphics[width=0.8 \linewidth]{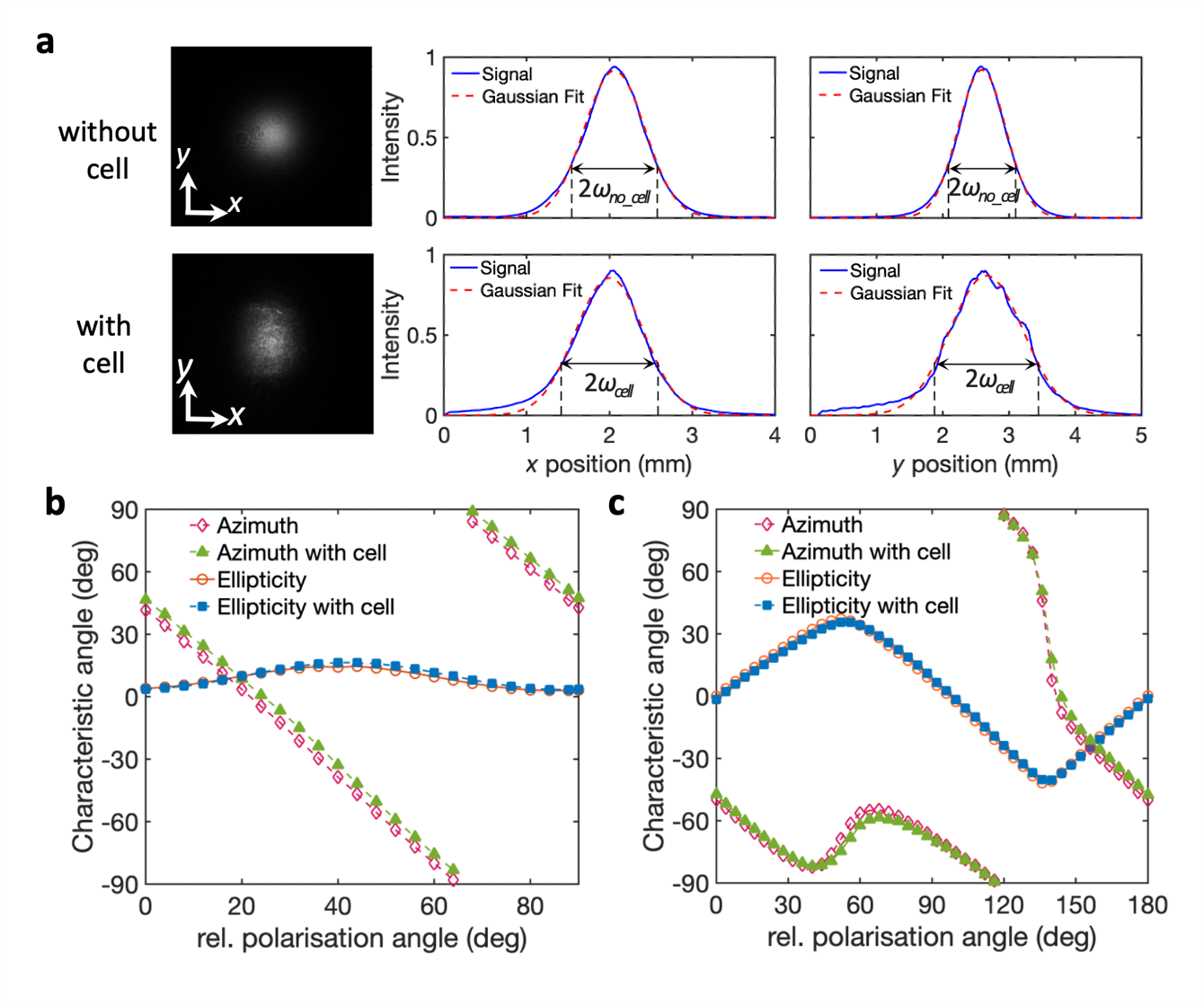}
\caption{\textbf{Transmission and polarisation measurement of the AM vapour cell.} \textbf{a} Beam profile of a 780 nm test laser beam with and without passing through the cell.\textbf{b} Azimuth and ellipticity measured using a polarisation analyser for varying angles of polarisation, with and without the presence of the printed vapour cell, for initially linearly polarised light. \textbf{c} The Azimuth and ellipticity for varying angles of a preceding quarter-waveplate, with and without the presence of the printed vapour cell, for initially linearly polarised light.}
\label{fig:Optical}
\end{figure*}

%%%%%%% Fig. 5

\begin{figure*}[ht]
\centering
\includegraphics[width=0.7 \linewidth]{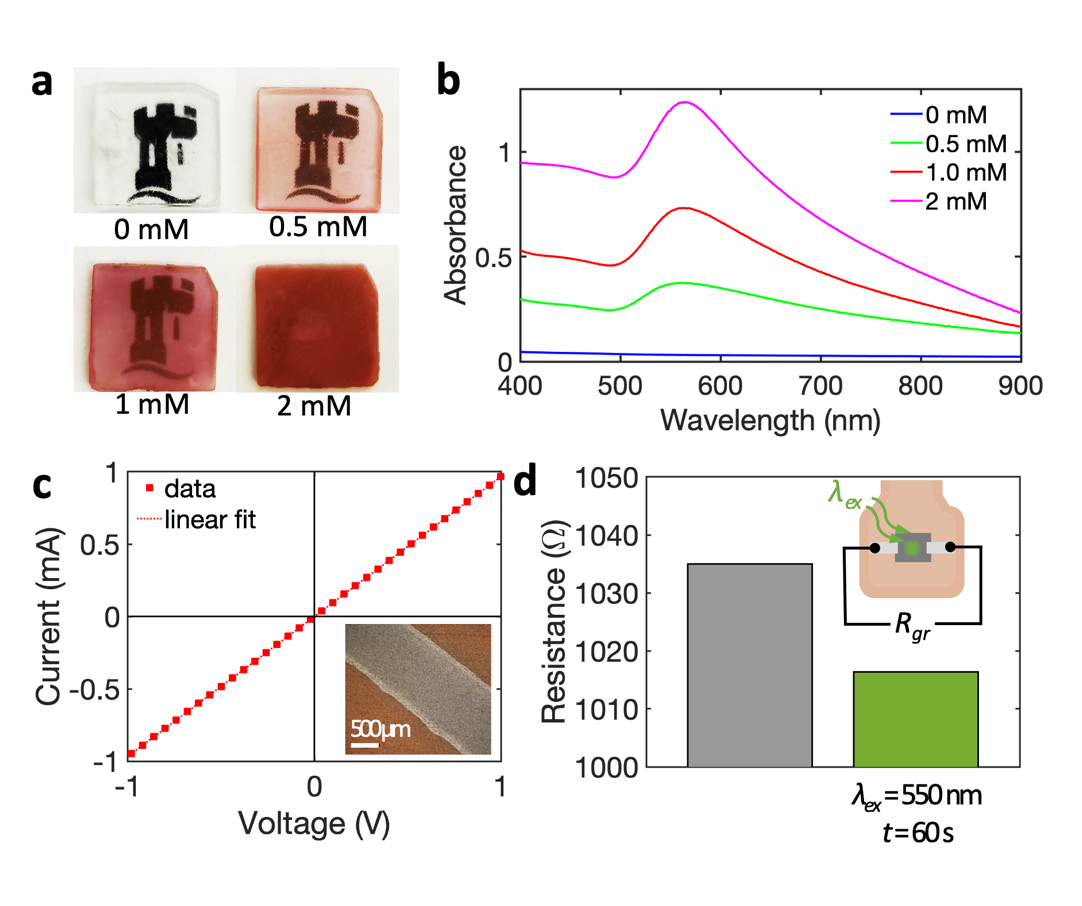}
\caption{\textbf{Functionalisation of AM glass - AuNP doping and graphene tracks.}  \textbf{a} Photographs of printed glass slabs of $1.25 \times 1.25$\,cm, with different concentrations of AuNP dopant, positioned over the University of Nottingham logo. \textbf{b} Measured absorbance of glass printed with different concentrations of AuNP dopant. \textbf{c} Representative I(V) dependence of a graphene track (10 mm by 2 mm) inkjet printed onto the 0.5 mM AuNP doped glass and (inset) optical microscopy image of the printed graphene track. \textbf{d} Electrical resistance, Rgr, of the graphene track printed onto an AuNP doped glass slab (0.5mM) measured in the dark and under continuous illumination with green laser light ($\lambda = 550$\,nm, $P = 25$\,mW). Inset: Schematic of the measurement scheme.}
\label{fig:cellfabrication}
\end{figure*}

Doppler-free spectroscopy enables laser frequency stabilisation (“locking”) to the hyperfine transitions. We stabilised a laser to the $^{85}$Rb $F = 3 \rightarrow F' = 3 \times 4$ crossover transition and thus demonstrated that the cell can be used as a frequency standard. The laser current was modulated with a frequency of 100 kHz, the photodiode signal was demodulated and an error signal obtained that corresponds to the derivative of the spectroscopy signal. This signal was used to frequency-stabilise the laser through feedback to the laser diode drive current. 

To analyse the resulting frequency stability, the error signal was recorded over a period of twenty minutes, while the laser was frequency locked via the signal from the printed vapour cell. By analysing the error signal, the Allan deviation was obtained for the locked laser and compared to the unlocked or free-running laser (see Figure \ref{fig:Allan_deviation}). 

The stabilisation based on the printed vapour cell reaches $\Delta F/F = 2 \times 10^{-10}$ at long interrogation times ($t \sim 1\,s$), an improvement of 1-2 orders of magnitude in comparison with the free-running laser, and provides significant stability improvements on all time scales. This demonstrates that the printed vapour cell can provide stable, Doppler-free laser locking. A similar setup based on polarisation can be used to operate the cell as a magnetometer\,\cite{Mitchell_2022, Palacios2022}. We compared the locking performance to a standard 75 mm long commercial vapour cell (Thorlabs, made by conventional glass blowing); taking the 15 times difference in optical path length into account, the Allan deviation normalised to optical path length is similar or slightly better for the printed cell (Figure \ref{fig:S6}).

To evaluate the applicability of printed vapour cells for magnetometry and polarisation dependent QT applications, the effect of the printed cell on beam shape and the polarisation of laser light (780 nm) was investigated with an internal vacuum of $2 \times 10^{-9}$ mbar. By analysing the Gaussian beam profile with and without the cell in the optical path, negligible distortion and a slight increase of the beam waist from the horizontally printed faces as shown in Figure \ref{fig:Optical}a. For the vertically printed faces, the beam waist increased by $\sim50\%$ [$\omega_{\mathrm{no\_cell}} = (0.51 \pm 0.1)$\,mm and $\omega_{\mathrm{cell}} = (0.78 \pm 0.1)$\,mm] most likely due to curvature of the internal surface, which can be considered in future designs. Small undulations are caused by the layer-by-layer printing process.

The effect of the printed vapour cell on polarisation was analysed, for linearly polarised light, for various input polarisation angles obtained via tuning of a half-wave plate, with a measurement of the azimuth and ellipticity taken every $4^{\circ}$ using a polarisation analyser. The measurements with and without the printed vapour cell (Figure \ref{fig:Optical}b) revealed negligible change (comparable to the resolution of the instrument) of azimuth and ellipticity when the cell is present in the optical path. The half-wave plate was then switched for a quarter-wave plate and the same measurement technique was repeated, to study the effect of the cell on elliptically polarised light. The results are shown in Figure 5c.

\subsection{Functionalisation of AM vapour cell}
Our method of 3D-printing vapour cells offers exceptional opportunities for improved quantum technologies. We are able to demonstrate a range of desirable features for vapour cell performance that are only made possible through additive manufacturing. The geometric design freedom of AM enabled printing of two inter-connected cells with a variable length tube (two 4.5\,mm cuboid cells connected with a 3\,mm channel, Figure \ref{fig:1}d), thus offering increased optical depth, while by overprinting the cell with conductive materials (Figure \ref{fig:1}e) we demonstrate integration and compactification potential for sensors and detectors. Tuneable modification of the optical transparency of 3D printed glass has previously been achieved by dipping the brown part into solutions containing metal salts\,\cite{kotz2017three,Liu2018,tabellion2006}. 
We propose a facile method where AuNP are formed in situ to modify the glass absorption in the visible wavelength range.\\
Gold salt, AuCl$_{3}$, was added to the resin before printing to form Au nanoparticles (AuNPs) in situ by a photothermal reduction process, thereby causing the glass to acquire a cranberry-red colour\,\cite{Hu2017}.
An absorption peak, characteristic of AuNPs, was observed at a wavelength of $ \sim 565\,$nm (Figure \ref{fig:cellfabrication}b)  for the glass produced with 1 mM of AuCl$_3$, indicating the formation of AuNPs with a size of $\sim 20\,$nm. By varying the AuCl$_3$ concentration from 0 to 2\,mM, the colour of the resulting glass changes to dark deep red (Figure \ref{fig:cellfabrication}a), indicating the formation of larger size AuNPs, and the absorption peak shifts to a longer wavelength from 562\,nm for 0.5\,mM to 567\,nm for 2\,mM (Figure \ref{fig:cellfabrication}b). Importantly, the doped glass retains sufficient optical transparency in the near infrared range, as required for Rb atomic spectroscopy, while filtering out shorter wavelength visible light from environmental sources. In the context of magnetometer cells, doped glass offers the exciting exploitation of surface plasmon resonances (SPR) of AuNP for local optically-activated heating, which can be used to achieve higher vapour pressure in the cell faster than standard current-based heating methods\,\cite{bharadwaj2014gold}.

To demonstrate SPR, we inkjet printed a graphene (iGr) track (10 mm (L) × 2 mm (W)) onto one side of the glass slab as shown in Figure \ref{fig:cellfabrication}c. Upon exposure of the undecorated side of the AuNP-glass to green laser light ($\lambda=550$\,nm), the resistance of the graphene track decreased by 2\% in 60 s (Figure \ref{fig:cellfabrication}d), which is a typical change in resistance expected for a temperature increase of $30\,^{\circ}$C. An equivalent track on undoped glass produced no detectable response under the same conditions (Figure \ref{fig:S4}a), confirming that the localised heating was generated by SPR of AuNPs. For medical sensing applications, e.g. magnetometry for brain imaging, this is a very relevant feature as the vapour pressure can be increased without heating coils and without raising the temperature of the environment; this reduces power dissipation and the need for thermal insulation, hence allowing the sensor to be brought closer to the tissues being studied. The technique also enables glass with a customisable optical absorption range within one glass structure (see Figure \ref{fig:S4} b)). 

Similarly to the overprinted graphene path, the vapour cell can be surface-decorated with other functional materials, including conductive layers/devices that can be inkjet printed onto the cell. A pair of 300 µm wide interdigitated conductive tracks was deposited using inkjet printing of silver nanoparticle ink and graphene inks (Figure \ref{fig:1}e), with a sheet resistance of $0.53\,\Omega/$sq and $207\,\Omega/$sq respectively, which is comparable to the values achieved on other substrates\,\cite{trindade2021residual,wang2021inter}. The conductive tracks exhibit good adhesion to the printed glass cell, enabling e.g. heating or magnetic field shielding elements that are co-manufactured with, and integral to, the vapour cell. The AM build process  also facilitates integration of other active components, such as inkjet-printed perovskite photon sensors\,\cite{austin2023photosensitisation}, which can be used for in situ monitoring.

\section{Conclusions}
Additive manufacturing of transparent glass elements has immense potential to improve components for quantum technologies. A glass vapour cell has been produced by digital light processing and a wide range of functionalisation capabilities have been demonstrated, including integrated 3D printed electronics and active optoelectronic components for enhanced device functionality, and wavelength-selective absorption for surface plasmonic heating.

The printed vapour cell was successfully pumped to the UHV regime ($2 \times 10^{-9}$\,mbar) and loaded with atomic rubidium vapour. The cell exhibits high transparency and polarisation stability, enabling the observation of atomic spectroscopy, including signature Doppler-valley and sub-Doppler features. Used as a frequency reference, laser stabilisation has been achieved with an Allan deviation $\Delta F/F < 10^{-9}$, limited by electronic feedback hardware, not the AM vapour cell. 

This work will inspire new AM-QT research directions, for example, how the sensitivity of a spin-exchange relaxation-free magnetometer based on modified surfaces in a 3D printed vapour cell compares to established designs\,\cite{serf}. The process could enable manufacture of $> 40$ glass cells (green parts) per lab-based printer per hour and the times required for debinding and sintering each cell could be effectively shortened with large volume furnaces. 

Our methods open the way for the creation of novel, high-performance sensors based on functionalised glass vapour cells for enhanced quantum technology devices, such as magneto-encephalography sensors or atomic clocks.  The upscalability of the AM process, together with good optical quality and high vacuum performance, makes such devices a scalable, convenient and customisable fabrication solution for many QT application sectors.

\section*{Data availability statement}
The data used in this work are available from the corresponding author upon reasonable request. 

\section*{Acknowledgements}
The fabrication work of the glass vapour cell was supported by EPSRC grants EP/P031684/1, EP/T001046/1 and EP/M013294/1. The characterisation work was supported by IUK project No.133086, EPSRC grants EP/T001046/1, EP/R024111/1 and EP/M013294/1, and by the European Comission grant ErBeStA (no. 800942).

\section*{Experimental Methods}
\textbf{Resin formulation}: A mixture of 60\,vol\% 2-hydroxyethyl methacrylate (HEMA), 10\,vol\% tetra (ethylene glycol) diacrylate (TEGDA) and 30\,vol\% phenoxyethanol (POE) was prepared, and blended with 0.2\,wt\% of bis (2,4,6 trimethylbenzoyl)phenylphosphineoxide as a photoinitiator, 0.1\,wt\% of hydroquinone monomethyl ether as an inhibitor and 0.035\,wt\% of Sudan Orange UV absorber. Aerosil OX50 silica nanopowder (50\,wt\%) was added in 10 small doses with 15 min bath sonication to prevent agglomeration. The final slurry was further bath sonicated for 30\,mins and then degassed at a low pressure of 200\,mbar for 3\,min. \\

\textbf{Additive manufacturing process}: The design was printed on a Cellink Lumen X printer (45\,mW/cm$^2$ UV power intensity at 405\,nm) with $50\,\mu$m hatching distance and 6.5\,s exposure time for each layer. The printed part was immersed in propylene glycol methyl ether acetate (PGMEA) and washed using a tube roller at 80\,rpm for 5\,minutes to remove the residual non-polymerised material. The washing cycle was repeated three times. Then the part was exposed to ultraviolet floodlight using a Wicked Curebox for 10\,minutes for post-curing. \\

\textbf{Post-process debinding and sintering}: The green part was debound in a high-temperature furnace (Carbolite HRF 7/22 box furnace) with the ambient temperature being increased at 0.35$^\circ$C/min to $T = 130^{\circ}$C, held at this temperature for 2 hours, then increased at the same rate to $320^{\circ}$C for 4 hours and $600^{\circ}$C for 2 hours. The resulting brown part was further sintered at $T = 1150^{\circ}$C for 12 hours with $3^{\circ}$C/min heating and cooling rates under 1 bar argon flow using a tube furnace (Carbolite STF 15/50).\\

\textbf{Functionalisation}: Doping with Au nanoparticles was achieved by adding AuCl$_3$ to the resin, with final concentrations of 0.5\,mM, 1\,mM and 2\,mM, before adding fumed silica powder. Conductive tracks were deposited with a $300\,\mu$m wide interdigitated design using a Dimatix DMP-2831 inkjet printer and a Samba cartridgeof 2.4\,pL drop volume. Three layers of AgNP ink (XTPL IJ36) were deposited using a $20\,\mu$m drop spacing and ten layers of graphene ink (Merck 793663) were printed using $10\,\mu$m drop spacings.\\

\textbf{Electrical and optical characterisation}: The sheet resistance of the printed silver and graphene samples was measured using a Keithley 2400 sourcemeter and a micromanipulator with a four-probe method, also known as the Kelvin technique, to eliminate contact resistance. Each I-V curve was measured three times using both forward and backwards scanning. The derivative of the I-V curve reveals the conductivity of the printed line with the exact geometry measured by an optical microscope.\\

\textbf{Characterisation of  polarised light}: The laser beam was set up to pass through a linear polariser and a waveplate (either quarter-wave or half-wave as appropriate), before passing through the printed vapour cell, which was under internal vacuum so that no gases in the cell could alter the polarisation.  \\

\textbf{Atomic spectroscopy}: The cell was connected to a UHV-flange via an annealed copper tube and attached to the tube using UV-curing glue (Dymax OP-67-LS). The system shown in Fig. \ref{fig:spectroscopy}b was used to demonstrate the vacuum compatibility of the cells. Following bake-out for 24hrs at $150^{\circ}$C and leak-testing, pressures down to $2 \times 10^{-9}$\,mbar were achieved. Laser light resonant with the rubidium absorption lines from a Toptica 110 tapered amplifier laser was then guided into the cell and the resulting absorption measured on a Thorlabs DET10A amplified photodiode. Doppler-free spectroscopy was obtained using a pump-probe setup as shown in Fig. \ref{fig:spectroscopy}b with 0.06\,mW probe beam power and 0.3\,mW pump beam power in beams with a diameter of 1.25\,mm. The commercial reference vapour cell of \diameter$25.4 \mathrm{mm} \times 71.8$\,mm was purchased from Thorlabs (GC25075-RB). For laser locking with this commercial cell, the probe beam is retro-reflected back through the cell, instead of separate probe and pump beams.

%\bibliography{main_arxiv}
\input{main_arxiv.bbl}

\newpage

\onecolumngrid
\appendix

\section*{Appendix}

\renewcommand{\thesubsection}{A\arabic{subsection}} 

\subsection{Optimisation of printing parameters for the resin}

By comparing the curing depth of the resins containing different concentrations of UV absorbers of 0.01 wt\%, 0.02 wt\%, 0.035 wt\% and 0.05\%, as shown in Figure \ref{fig:S1}, the 0.035 wt\% was chosen as the optimum for the $50 \mu$m layer thickness used by the Lumen X printer, which results in a high geometrical fidelity at short printing times.

\renewcommand{\thefigure}{A\arabic{figure}}
\setcounter{figure}{0}

\begin{figure}[h]
\centering
\includegraphics[width=0.4 \columnwidth]{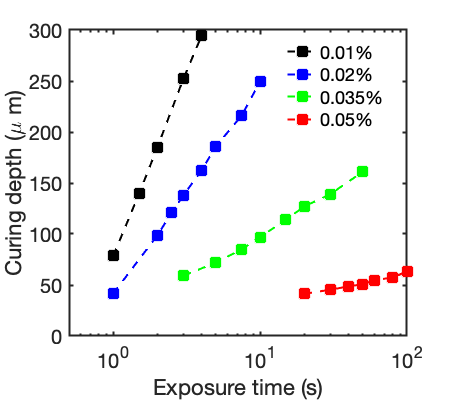}
\caption{Measured curing depth versus exposure time under \blue{irradiance of} 45\,mW/cm$^2$  UV light for resins with concentrations of UV absorbers of 0.01\%, 0.02\%, 0.035\% and 0.05\%.}
\label{fig:S1}
\end{figure}

\subsection{Modelling of scattered light of nanoparticles in the resin}
\begin{figure}[h]
\centering
\includegraphics[width=0.8 \columnwidth]{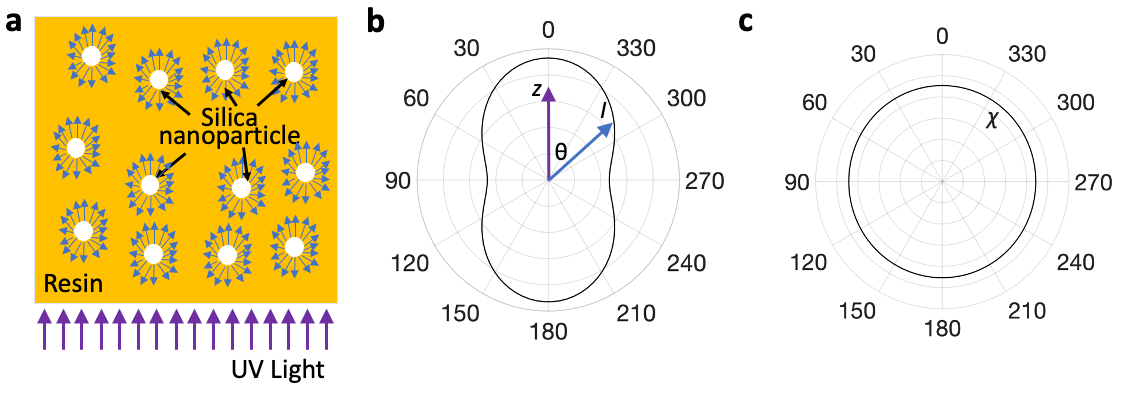}
\caption{a) Sketch showing the light scattering by the silica nanoparticles in the resin. b) Light scattering power density distribution from a silica nanoparticle when $R = 1.02 d$ from equ. \eqref{A1}. c) The resulting, homogeneous consumption of the vinyl group $\chi$ due to scattering around the nanoparticle from equ. \eqref{A2}, when $R = 1.02 d$ is chosen.}
\label{fig:S2}
\end{figure}

Previous studies have highlighted that the existence of nanoparticles in UV curable resins promotes the curing depth due to the light scattered from the nanoparticles\,\cite{Faria2017}, and the extra energy overcures the photocurable material in the curing layer which reduces the printing resolution or even interferes with the curing of the next layer in DLP printing (\ref{fig:S2}a)). Following previous work on the modelling of polymerisation kinetics\cite{Zhao2021}, the degree of curing caused by scattering in our resin formulation was calculated and confirmed that the formulated resin successfully suppressed the curing inhomogeneity caused by scattered light around the nanoparticles. 

The scattered light intensity distribution, $I(\theta)$, due to Rayleigh scattering of one OX 50 nanoparticle can be calculated by 
\begin{equation}
I(\theta) = I_0 \frac{1+\cos \theta}{2 R^2}\left(\frac{2\pi}{\lambda}\right)^4\left(\frac{n^2 -1}{n^2+2}\right)^2 \left(\frac{d}{2} \right)^6       \tag{A1}
\label{A1}
\end{equation}
where $I_{0} = 45$\,mW/cm$^2$ is the intensity of the Lumen X printer used in this work, $R$ is the mean distance between the particles, $\theta$ is the scattering angle, $n = 1.45$ is the refractive index of the resin, $d = 40$\,nm is the average diameter of the particle and $\lambda = 405\,$nm is the wavelength of the light.\\
Taking $R = 1.02\,d$, which matches the 50\,wt\% of nanoparticles in the resin, the distribution of the scattered light  intensity $I(\theta)$ is anistropic and the scattered light is 1.9 times stronger along the direction of the incoming UV light (z-direction) than the perpendicular x-y-plane (Figure \ref{fig:S1}b)). 

Despite the anistropic angular distribution of the scattered light intensity, the formulated resin exhibits an isotropic degree of polymerisation. The total consumption of the vinyl groups during polymerisation, $\chi$, can be described by a model developed in\,\cite{Zhao2021}, which combines classical theory of free radical polymerisation with consumption $\chi_p$, and an autoacceleration effect with consumption $\chi_a$,
\begin{equation}
\chi = \chi_{p}+\chi_{a}=\chi_{\mathrm{max,p}}\left(1-e^{k_p \Lambda} \right) +\frac{\chi_{\mathrm{max,a}}}{1+e^{k_a (\Lambda-\Lambda_C)}},     \tag{A2}
\label{A2}
\end{equation}
where $\Lambda=\sqrt{I}t$ is the UV radiation dose ($t$ = 6.5 s is the exposure time), $\chi_{\mathrm{max,p}} =0.2079$ and $\chi_{\mathrm{max,a}} = 0.5729$ are the maximum degrees of vinyl group consumption that can be achieved by the classical radical polymerisation and the autoacceleration polymerisation respectively. The rate constant of free radical polymerisation $k_p= -0.0083\mathrm{(kg/s)^{-1/2}}$, the rate constant for the autoacceleration stage $k_a = -0.1258\mathrm{(kg/s)^{-1/2}}$ and the required UV radiation dose to achieve half of the maximum degree of vinyl group consumption at the autoacceleration stage $\Lambda_C = 14.0719 \mathrm{(kg/s)^{1/2}}$ are parameters determined by the compositions of the monomers, photoinitiators and absorbers in the formulation utilising the method described in \cite{Zhao2021}. For the used resin formulation, the expected angular distribution for the degree of curing $\chi$ has an isotropic distribution as shown in Fig.\ref{fig:S2}, which indicates a similar degree of curing along all directions.

\newpage

\subsection{Characterisation of printed and functionalised glass structures}

\begin{figure}[h]
\centering
\includegraphics[width=0.8 \columnwidth]{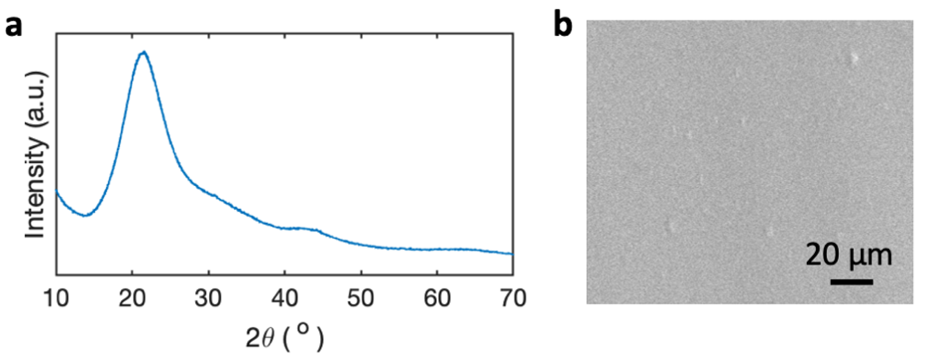}
\caption{a) X-ray diffraction (XRD) pattern and b) a representative, cross-section scanning electron microscope (SEM) image of the final printed glass part showing an amorphous structure after sintering.}
\label{fig:S3}
\end{figure}

\begin{figure}[h]
\centering
\includegraphics[width=0.7 \columnwidth]{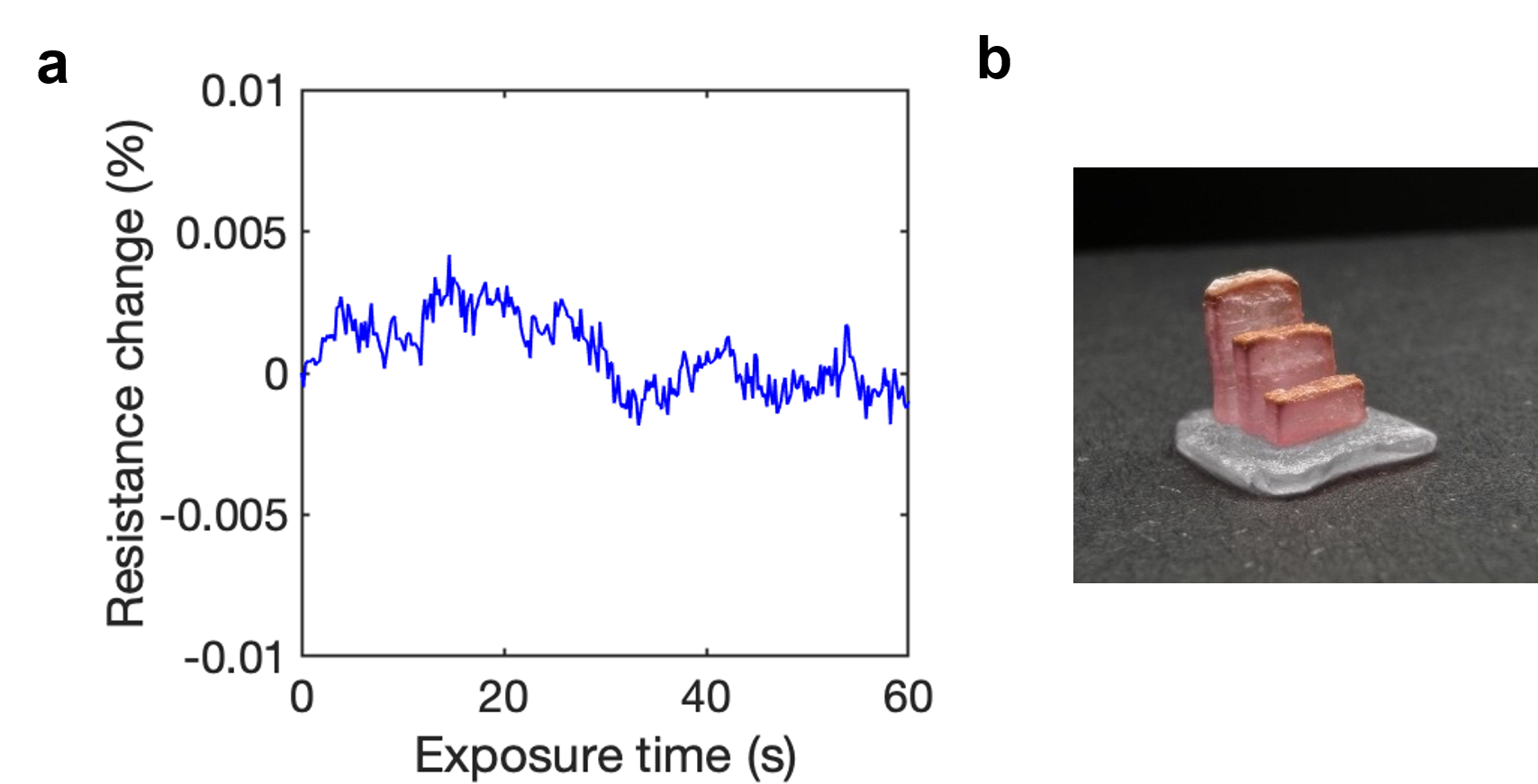}
\caption{a) Resistance change of a graphene track printed onto a non-doped glass slab during exposure to continuous illumination with green laser light (550\,nm, P = 25\,mW). b) Printed glass structures with two different colours. The base is without doping and the standing surfaces are doped with 0.5 mM AuNP.}
\label{fig:S4}
\end{figure}

%\vspace{4cm}

\subsection{Laser locking comparison to Thorlabs 75 mm vapour cell}
\begin{figure}[h]
\centering
\includegraphics[width=0.5 \columnwidth]{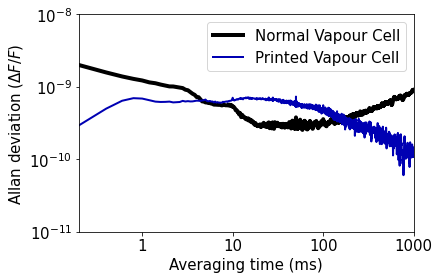}
\caption{Allan deviation comparison of the printed vapour cell and the commercial vapour cell, adjusted for the ratio of the corresponding optical path lengths.}
\label{fig:S6}
\end{figure}

For laser frequency stabilisation, we compared the performance of our printed vapour cell to a standard Thorlabs 75 mm vapour cell. For laser locking with this cell, the probe beam is retro-reflected back through the cell, instead of separate probe and pump beams. Note that the total path length is 15 times that for the printed vapour cell. The resulting Allen deviation is shown in Figure \ref{fig:S6}.

\vspace{1cm}

\newpage

\subsection{Polarisation characterisation of the printed vapour cell}
\begin{figure}[h]
\centering
\includegraphics[width=0.5 \columnwidth]{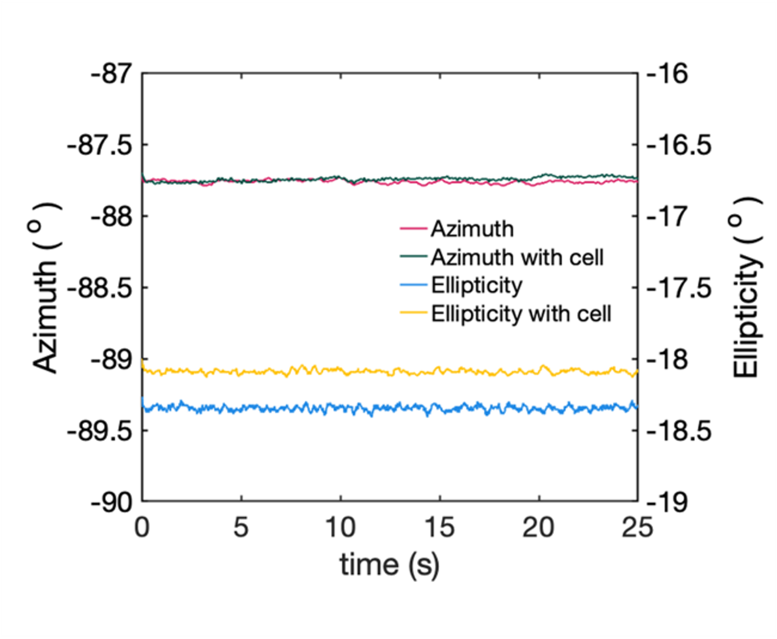}
\caption{The azimuth and ellipticity measured on a polarisation analyser over time, with and without the presence of the printed vapour cell, for a given incident polarisation.}
\label{fig:S7}
\end{figure}

The stability of the polarisation was also measured by analysing the polarisation over time (Figure \ref{fig:S7}). No time-dependent change (e.g. through a local temperature change) in polarisation was detected. \\

\vspace{2.5cm}

\end{document}

%% file: main_arxiv.bbl

%% file: main_arxiv.bbl
\begin{thebibliography}{59}%
\makeatletter
\providecommand \@ifxundefined [1]{%
 \@ifx{#1\undefined}
}%
\providecommand \@ifnum [1]{%
 \ifnum #1\expandafter \@firstoftwo
 \else \expandafter \@secondoftwo
 \fi
}%
\providecommand \@ifx [1]{%
 \ifx #1\expandafter \@firstoftwo
 \else \expandafter \@secondoftwo
 \fi
}%
\providecommand \natexlab [1]{#1}%
\providecommand \enquote  [1]{``#1''}%
\providecommand \bibnamefont  [1]{#1}%
\providecommand \bibfnamefont [1]{#1}%
\providecommand \citenamefont [1]{#1}%
\providecommand \href@noop [0]{\@secondoftwo}%
\providecommand \href [0]{\begingroup \@sanitize@url \@href}%
\providecommand \@href[1]{\@@startlink{#1}\@@href}%
\providecommand \@@href[1]{\endgroup#1\@@endlink}%
\providecommand \@sanitize@url [0]{\catcode `\\12\catcode `\$12\catcode `\&12\catcode `\#12\catcode `\^12\catcode `\_12\catcode `\%12\relax}%
\providecommand \@@startlink[1]{}%
\providecommand \@@endlink[0]{}%
\providecommand \url  [0]{\begingroup\@sanitize@url \@url }%
\providecommand \@url [1]{\endgroup\@href {#1}{\urlprefix }}%
\providecommand \urlprefix  [0]{URL }%
\providecommand \Eprint [0]{\href }%
\providecommand \doibase [0]{https://doi.org/}%
\providecommand \selectlanguage [0]{\@gobble}%
\providecommand \bibinfo  [0]{\@secondoftwo}%
\providecommand \bibfield  [0]{\@secondoftwo}%
\providecommand \translation [1]{[#1]}%
\providecommand \BibitemOpen [0]{}%
\providecommand \bibitemStop [0]{}%
\providecommand \bibitemNoStop [0]{.\EOS\space}%
\providecommand \EOS [0]{\spacefactor3000\relax}%
\providecommand \BibitemShut  [1]{\csname bibitem#1\endcsname}%
\let\auto@bib@innerbib\@empty
%</preamble>
\bibitem [{\citenamefont {Aslam}\ \emph {et~al.}(2023)\citenamefont {Aslam}, \citenamefont {Zhou}, \citenamefont {Urbach}, \citenamefont {Turner}, \citenamefont {Walsworth}, \citenamefont {Lukin},\ and\ \citenamefont {Park}}]{aslam2023quantum}%
  \BibitemOpen
  \bibfield  {author} {\bibinfo {author} {\bibfnamefont {N.}~\bibnamefont {Aslam}}, \bibinfo {author} {\bibfnamefont {H.}~\bibnamefont {Zhou}}, \bibinfo {author} {\bibfnamefont {E.~K.}\ \bibnamefont {Urbach}}, \bibinfo {author} {\bibfnamefont {M.~J.}\ \bibnamefont {Turner}}, \bibinfo {author} {\bibfnamefont {R.~L.}\ \bibnamefont {Walsworth}}, \bibinfo {author} {\bibfnamefont {M.~D.}\ \bibnamefont {Lukin}},\ and\ \bibinfo {author} {\bibfnamefont {H.}~\bibnamefont {Park}},\ }\bibfield  {title} {\bibinfo {title} {Quantum sensors for biomedical applications},\ }\href {https://doi.org/https://doi.org/10.1038/s42254-023-00558-3} {\bibfield  {journal} {\bibinfo  {journal} {Nature Reviews Physics}\ }\textbf {\bibinfo {volume} {5}},\ \bibinfo {pages} {157} (\bibinfo {year} {2023})}\BibitemShut {NoStop}%
\bibitem [{\citenamefont {Boto}\ \emph {et~al.}(2018)\citenamefont {Boto}, \citenamefont {Holmes}, \citenamefont {Legget}, \citenamefont {Roberts}, \citenamefont {Shah}, \citenamefont {Meyer}, \citenamefont {Munoz}, \citenamefont {Mullinger}, \citenamefont {Tierney}, \citenamefont {Bestmann}, \citenamefont {Barnes}, \citenamefont {Bowtell},\ and\ \citenamefont {Brookes}}]{SPMIC}%
  \BibitemOpen
  \bibfield  {author} {\bibinfo {author} {\bibfnamefont {E.}~\bibnamefont {Boto}}, \bibinfo {author} {\bibfnamefont {N.}~\bibnamefont {Holmes}}, \bibinfo {author} {\bibfnamefont {J.}~\bibnamefont {Legget}}, \bibinfo {author} {\bibfnamefont {G.}~\bibnamefont {Roberts}}, \bibinfo {author} {\bibfnamefont {V.}~\bibnamefont {Shah}}, \bibinfo {author} {\bibfnamefont {S.}~\bibnamefont {Meyer}}, \bibinfo {author} {\bibfnamefont {L.}~\bibnamefont {Munoz}}, \bibinfo {author} {\bibfnamefont {K.}~\bibnamefont {Mullinger}}, \bibinfo {author} {\bibfnamefont {T.}~\bibnamefont {Tierney}}, \bibinfo {author} {\bibfnamefont {S.}~\bibnamefont {Bestmann}}, \bibinfo {author} {\bibfnamefont {G.}~\bibnamefont {Barnes}}, \bibinfo {author} {\bibfnamefont {R.}~\bibnamefont {Bowtell}},\ and\ \bibinfo {author} {\bibfnamefont {M.}~\bibnamefont {Brookes}},\ }\bibfield  {title} {\bibinfo {title} {Moving magnetoencephalography towards real-world applications with a wearable system},\ }\href
  {https://doi.org/https://doi.org/10.1038/nature26147} {\bibfield  {journal} {\bibinfo  {journal} {Nature}\ }\textbf {\bibinfo {volume} {555}},\ \bibinfo {pages} {657} (\bibinfo {year} {2018})}\BibitemShut {NoStop}%
\bibitem [{\citenamefont {Arita}\ \emph {et~al.}(2013)\citenamefont {Arita}, \citenamefont {Mazilu},\ and\ \citenamefont {Dholakia}}]{arita2013laser}%
  \BibitemOpen
  \bibfield  {author} {\bibinfo {author} {\bibfnamefont {Y.}~\bibnamefont {Arita}}, \bibinfo {author} {\bibfnamefont {M.}~\bibnamefont {Mazilu}},\ and\ \bibinfo {author} {\bibfnamefont {K.}~\bibnamefont {Dholakia}},\ }\bibfield  {title} {\bibinfo {title} {Laser-induced rotation and cooling of a trapped microgyroscope in vacuum},\ }\href {https://doi.org/https://doi.org/10.1038/ncomms3374} {\bibfield  {journal} {\bibinfo  {journal} {Nature communications}\ }\textbf {\bibinfo {volume} {4}},\ \bibinfo {pages} {2374} (\bibinfo {year} {2013})}\BibitemShut {NoStop}%
\bibitem [{\citenamefont {Bongs}\ \emph {et~al.}(2019)\citenamefont {Bongs}, \citenamefont {Holynski}, \citenamefont {Vovrosh}, \citenamefont {Bouyer}, \citenamefont {Condon}, \citenamefont {Rasel}, \citenamefont {Schubert}, \citenamefont {Schleich},\ and\ \citenamefont {Roura}}]{Bongs2019}%
  \BibitemOpen
  \bibfield  {author} {\bibinfo {author} {\bibfnamefont {K.}~\bibnamefont {Bongs}}, \bibinfo {author} {\bibfnamefont {M.}~\bibnamefont {Holynski}}, \bibinfo {author} {\bibfnamefont {J.}~\bibnamefont {Vovrosh}}, \bibinfo {author} {\bibfnamefont {P.}~\bibnamefont {Bouyer}}, \bibinfo {author} {\bibfnamefont {G.}~\bibnamefont {Condon}}, \bibinfo {author} {\bibfnamefont {E.}~\bibnamefont {Rasel}}, \bibinfo {author} {\bibfnamefont {C.}~\bibnamefont {Schubert}}, \bibinfo {author} {\bibfnamefont {W.}~\bibnamefont {Schleich}},\ and\ \bibinfo {author} {\bibfnamefont {A.}~\bibnamefont {Roura}},\ }\href {https://doi.org/https://doi.org/10.1038/s41586-021-04315-3} {\bibfield  {journal} {\bibinfo  {journal} {Nature Reviews Physics}\ }\textbf {\bibinfo {volume} {1}},\ \bibinfo {pages} {731} (\bibinfo {year} {2019})}\BibitemShut {NoStop}%
\bibitem [{\citenamefont {Lee}\ \emph {et~al.}(2023)\citenamefont {Lee}, \citenamefont {Lisanti}, \citenamefont {Terrano},\ and\ \citenamefont {Romalis}}]{Lee2023}%
  \BibitemOpen
  \bibfield  {author} {\bibinfo {author} {\bibfnamefont {J.}~\bibnamefont {Lee}}, \bibinfo {author} {\bibfnamefont {M.}~\bibnamefont {Lisanti}}, \bibinfo {author} {\bibfnamefont {W.~A.}\ \bibnamefont {Terrano}},\ and\ \bibinfo {author} {\bibfnamefont {M.}~\bibnamefont {Romalis}},\ }\bibfield  {title} {\bibinfo {title} {Laboratory constraints on the neutron-spin coupling of fev-scale axions},\ }\href {https://doi.org/10.1103/PhysRevX.13.011050} {\bibfield  {journal} {\bibinfo  {journal} {Phys. Rev. X}\ }\textbf {\bibinfo {volume} {13}},\ \bibinfo {pages} {011050} (\bibinfo {year} {2023})}\BibitemShut {NoStop}%
\bibitem [{\citenamefont {Feng}(2019)}]{feng2019review}%
  \BibitemOpen
  \bibfield  {author} {\bibinfo {author} {\bibfnamefont {D.}~\bibnamefont {Feng}},\ }\bibfield  {title} {\bibinfo {title} {Review of quantum navigation},\ }in\ \href@noop {} {\emph {\bibinfo {booktitle} {IOP Conference Series: Earth and Environmental Science}}},\ Vol.\ \bibinfo {volume} {237}\ (\bibinfo {organization} {IOP Publishing},\ \bibinfo {year} {2019})\ p.\ \bibinfo {pages} {032027}\BibitemShut {NoStop}%
\bibitem [{\citenamefont {Stray}\ \emph {et~al.}(2022)\citenamefont {Stray}, \citenamefont {Lamb}, \citenamefont {Kaushik}, \citenamefont {Vovrosh}, \citenamefont {Rodgers}, \citenamefont {Winch}, \citenamefont {Hayati}, \citenamefont {Boddice}, \citenamefont {Strabawa},\ and\ \citenamefont {Niggebaum}}]{Stray2022}%
  \BibitemOpen
  \bibfield  {author} {\bibinfo {author} {\bibfnamefont {B.}~\bibnamefont {Stray}}, \bibinfo {author} {\bibfnamefont {A.}~\bibnamefont {Lamb}}, \bibinfo {author} {\bibfnamefont {A.}~\bibnamefont {Kaushik}}, \bibinfo {author} {\bibfnamefont {J.}~\bibnamefont {Vovrosh}}, \bibinfo {author} {\bibfnamefont {A.}~\bibnamefont {Rodgers}}, \bibinfo {author} {\bibfnamefont {J.}~\bibnamefont {Winch}}, \bibinfo {author} {\bibfnamefont {F.}~\bibnamefont {Hayati}}, \bibinfo {author} {\bibfnamefont {D.}~\bibnamefont {Boddice}}, \bibinfo {author} {\bibfnamefont {A.}~\bibnamefont {Strabawa}},\ and\ \bibinfo {author} {\bibfnamefont {A.}~\bibnamefont {Niggebaum}},\ }\bibfield  {title} {\bibinfo {title} {Quantum sensing for gravity cartography},\ }\href {https://doi.org/https://doi.org/10.1038/s41586-021-04315-3} {\bibfield  {journal} {\bibinfo  {journal} {Nature}\ }\textbf {\bibinfo {volume} {602}},\ \bibinfo {pages} {590} (\bibinfo {year} {2022})}\BibitemShut {NoStop}%
\bibitem [{\citenamefont {Evered}\ \emph {et~al.}(2023)\citenamefont {Evered}, \citenamefont {Bluvstein}, \citenamefont {Kalinowski}, \citenamefont {Ebadi}, \citenamefont {Manovitz}, \citenamefont {Zhou}, \citenamefont {Li}, \citenamefont {Geim}, \citenamefont {Wang}, \citenamefont {Maskara}, \citenamefont {Levine}, \citenamefont {Semeghini}, \citenamefont {Greiner}, \citenamefont {Vuletic},\ and\ \citenamefont {Lukin}}]{Evered2023}%
  \BibitemOpen
  \bibfield  {author} {\bibinfo {author} {\bibfnamefont {S.~J.}\ \bibnamefont {Evered}}, \bibinfo {author} {\bibfnamefont {D.}~\bibnamefont {Bluvstein}}, \bibinfo {author} {\bibfnamefont {M.}~\bibnamefont {Kalinowski}}, \bibinfo {author} {\bibfnamefont {S.}~\bibnamefont {Ebadi}}, \bibinfo {author} {\bibfnamefont {T.}~\bibnamefont {Manovitz}}, \bibinfo {author} {\bibfnamefont {H.}~\bibnamefont {Zhou}}, \bibinfo {author} {\bibfnamefont {S.~H.}\ \bibnamefont {Li}}, \bibinfo {author} {\bibfnamefont {A.~A.}\ \bibnamefont {Geim}}, \bibinfo {author} {\bibfnamefont {T.~T.}\ \bibnamefont {Wang}}, \bibinfo {author} {\bibfnamefont {N.}~\bibnamefont {Maskara}}, \bibinfo {author} {\bibfnamefont {H.}~\bibnamefont {Levine}}, \bibinfo {author} {\bibfnamefont {G.}~\bibnamefont {Semeghini}}, \bibinfo {author} {\bibfnamefont {M.}~\bibnamefont {Greiner}}, \bibinfo {author} {\bibfnamefont {V.}~\bibnamefont {Vuletic}},\ and\ \bibinfo {author} {\bibfnamefont {M.~D.}\ \bibnamefont {Lukin}},\ }\bibfield  {title} {\bibinfo {title}
  {High-fidelity parallel entangling gates on a neutral-atom quantum computer},\ }\href {https://doi.org/https://doi.org/10.1038/s41586-023-06481-y} {\bibfield  {journal} {\bibinfo  {journal} {Nature}\ }\textbf {\bibinfo {volume} {622}},\ \bibinfo {pages} {268} (\bibinfo {year} {2023})}\BibitemShut {NoStop}%
\bibitem [{\citenamefont {Ruchka}\ \emph {et~al.}(2022)\citenamefont {Ruchka}, \citenamefont {Hammer}, \citenamefont {Rockenhäuser}, \citenamefont {Albrecht}, \citenamefont {Drozella}, \citenamefont {Thiele}, \citenamefont {Giessen},\ and\ \citenamefont {Langen}}]{Ruchka2022}%
  \BibitemOpen
  \bibfield  {author} {\bibinfo {author} {\bibfnamefont {P.}~\bibnamefont {Ruchka}}, \bibinfo {author} {\bibfnamefont {S.}~\bibnamefont {Hammer}}, \bibinfo {author} {\bibfnamefont {M.}~\bibnamefont {Rockenhäuser}}, \bibinfo {author} {\bibfnamefont {R.}~\bibnamefont {Albrecht}}, \bibinfo {author} {\bibfnamefont {J.}~\bibnamefont {Drozella}}, \bibinfo {author} {\bibfnamefont {S.}~\bibnamefont {Thiele}}, \bibinfo {author} {\bibfnamefont {H.}~\bibnamefont {Giessen}},\ and\ \bibinfo {author} {\bibfnamefont {T.}~\bibnamefont {Langen}},\ }\bibfield  {title} {\bibinfo {title} {Microscopic 3d printed optical tweezers for atomic quantum technology},\ }\href {https://doi.org/DOI 10.1088/2058-9565/ac796c} {\bibfield  {journal} {\bibinfo  {journal} {Quantum Science and Technology}\ }\textbf {\bibinfo {volume} {7}},\ \bibinfo {pages} {045011} (\bibinfo {year} {2022})}\BibitemShut {NoStop}%
\bibitem [{\citenamefont {Vovrosh}\ \emph {et~al.}(2018)\citenamefont {Vovrosh}, \citenamefont {Voulazeris}, \citenamefont {Petrov}, \citenamefont {Zou}, \citenamefont {Gaber}, \citenamefont {Benn}, \citenamefont {Woolger}, \citenamefont {Attallah}, \citenamefont {Boyer}, \citenamefont {Bongs},\ and\ \citenamefont {Holynski}}]{vovrosh2018}%
  \BibitemOpen
  \bibfield  {author} {\bibinfo {author} {\bibfnamefont {J.}~\bibnamefont {Vovrosh}}, \bibinfo {author} {\bibfnamefont {G.}~\bibnamefont {Voulazeris}}, \bibinfo {author} {\bibfnamefont {P.}~\bibnamefont {Petrov}}, \bibinfo {author} {\bibfnamefont {J.}~\bibnamefont {Zou}}, \bibinfo {author} {\bibfnamefont {Y.}~\bibnamefont {Gaber}}, \bibinfo {author} {\bibfnamefont {L.}~\bibnamefont {Benn}}, \bibinfo {author} {\bibfnamefont {D.}~\bibnamefont {Woolger}}, \bibinfo {author} {\bibfnamefont {M.~M.}\ \bibnamefont {Attallah}}, \bibinfo {author} {\bibfnamefont {V.}~\bibnamefont {Boyer}}, \bibinfo {author} {\bibfnamefont {K.}~\bibnamefont {Bongs}},\ and\ \bibinfo {author} {\bibfnamefont {M.}~\bibnamefont {Holynski}},\ }\bibfield  {title} {\bibinfo {title} {Additive manufacturing of magnetic shielding and ultra-high vacuum flange for cold atom sensors},\ }\href {https://doi.org/https://doi.org/10.1038/s41598-018-20352-x} {\bibfield  {journal} {\bibinfo  {journal} {Scientific Reports}\ }\textbf {\bibinfo {volume} {8}},\
  \bibinfo {pages} {2023} (\bibinfo {year} {2018})}\BibitemShut {NoStop}%
\bibitem [{\citenamefont {Cooper}\ \emph {et~al.}(2021)\citenamefont {Cooper}, \citenamefont {Coles}, \citenamefont {Everton}, \citenamefont {Makery}, \citenamefont {Campion}, \citenamefont {Madkhaly}, \citenamefont {Morley}, \citenamefont {O'Shea}, \citenamefont {Evans}, \citenamefont {Saint}, \citenamefont {Kr\"{u}ger}, \citenamefont {Orucevic}, \citenamefont {Tuck}, \citenamefont {Wildman}, \citenamefont {Fromhold},\ and\ \citenamefont {Hackerm\"{u}ller}}]{cooper2021printedchamber}%
  \BibitemOpen
  \bibfield  {author} {\bibinfo {author} {\bibfnamefont {N.}~\bibnamefont {Cooper}}, \bibinfo {author} {\bibfnamefont {L.}~\bibnamefont {Coles}}, \bibinfo {author} {\bibfnamefont {S.}~\bibnamefont {Everton}}, \bibinfo {author} {\bibfnamefont {I.}~\bibnamefont {Makery}}, \bibinfo {author} {\bibfnamefont {R.}~\bibnamefont {Campion}}, \bibinfo {author} {\bibfnamefont {S.}~\bibnamefont {Madkhaly}}, \bibinfo {author} {\bibfnamefont {C.}~\bibnamefont {Morley}}, \bibinfo {author} {\bibfnamefont {J.}~\bibnamefont {O'Shea}}, \bibinfo {author} {\bibfnamefont {W.}~\bibnamefont {Evans}}, \bibinfo {author} {\bibfnamefont {R.}~\bibnamefont {Saint}}, \bibinfo {author} {\bibfnamefont {P.}~\bibnamefont {Kr\"{u}ger}}, \bibinfo {author} {\bibfnamefont {F.}~\bibnamefont {Orucevic}}, \bibinfo {author} {\bibfnamefont {C.}~\bibnamefont {Tuck}}, \bibinfo {author} {\bibfnamefont {R.}~\bibnamefont {Wildman}}, \bibinfo {author} {\bibfnamefont {T.}~\bibnamefont {Fromhold}},\ and\ \bibinfo {author} {\bibfnamefont {L.}~\bibnamefont
  {Hackerm\"{u}ller}},\ }\bibfield  {title} {\bibinfo {title} {Additively manufactured ultra-high vacuum chamber for portable quantum technologies},\ }\href {https://doi.org/https://doi.org/10.1016/j.addma.2021.101898} {\bibfield  {journal} {\bibinfo  {journal} {Additive Manufacturinig}\ }\textbf {\bibinfo {volume} {40}},\ \bibinfo {pages} {101898} (\bibinfo {year} {2021})}\BibitemShut {NoStop}%
\bibitem [{\citenamefont {Madkhaly}\ \emph {et~al.}(2021)\citenamefont {Madkhaly}, \citenamefont {Coles}, \citenamefont {Morley}, \citenamefont {Colquhoun}, \citenamefont {Fromhold}, \citenamefont {Cooper},\ and\ \citenamefont {Hackerm\"uller}}]{madkhaly2021printedoptics}%
  \BibitemOpen
  \bibfield  {author} {\bibinfo {author} {\bibfnamefont {S.}~\bibnamefont {Madkhaly}}, \bibinfo {author} {\bibfnamefont {L.}~\bibnamefont {Coles}}, \bibinfo {author} {\bibfnamefont {C.}~\bibnamefont {Morley}}, \bibinfo {author} {\bibfnamefont {C.}~\bibnamefont {Colquhoun}}, \bibinfo {author} {\bibfnamefont {T.}~\bibnamefont {Fromhold}}, \bibinfo {author} {\bibfnamefont {N.}~\bibnamefont {Cooper}},\ and\ \bibinfo {author} {\bibfnamefont {L.}~\bibnamefont {Hackerm\"uller}},\ }\bibfield  {title} {\bibinfo {title} {Performance-optimized components for quantum technologies via additive manufacturing},\ }\href {https://doi.org/10.1103/PRXQuantum.2.030326} {\bibfield  {journal} {\bibinfo  {journal} {PRX Quantum}\ }\textbf {\bibinfo {volume} {2}},\ \bibinfo {pages} {030326} (\bibinfo {year} {2021})}\BibitemShut {NoStop}%
\bibitem [{\citenamefont {St\ae{}rkind}\ \emph {et~al.}(2023)\citenamefont {St\ae{}rkind}, \citenamefont {Jensen}, \citenamefont {M\"uller}, \citenamefont {Boer}, \citenamefont {Petersen},\ and\ \citenamefont {Polzik}}]{Staerkind2023}%
  \BibitemOpen
  \bibfield  {author} {\bibinfo {author} {\bibfnamefont {H.}~\bibnamefont {St\ae{}rkind}}, \bibinfo {author} {\bibfnamefont {K.}~\bibnamefont {Jensen}}, \bibinfo {author} {\bibfnamefont {J.~H.}\ \bibnamefont {M\"uller}}, \bibinfo {author} {\bibfnamefont {V.~O.}\ \bibnamefont {Boer}}, \bibinfo {author} {\bibfnamefont {E.~T.}\ \bibnamefont {Petersen}},\ and\ \bibinfo {author} {\bibfnamefont {E.~S.}\ \bibnamefont {Polzik}},\ }\bibfield  {title} {\bibinfo {title} {Precision measurement of the excited state land\'e g-factor and diamagnetic shift of the cesium ${\mathrm{d}}_{2}$ line},\ }\href {https://doi.org/10.1103/PhysRevX.13.021036} {\bibfield  {journal} {\bibinfo  {journal} {Phys. Rev. X}\ }\textbf {\bibinfo {volume} {13}},\ \bibinfo {pages} {021036} (\bibinfo {year} {2023})}\BibitemShut {NoStop}%
\bibitem [{\citenamefont {Liew}\ \emph {et~al.}(2004)\citenamefont {Liew}, \citenamefont {Knappe}, \citenamefont {Moreland}, \citenamefont {Robinson}, \citenamefont {Hollberg},\ and\ \citenamefont {Kitching}}]{Liew2004}%
  \BibitemOpen
  \bibfield  {author} {\bibinfo {author} {\bibfnamefont {L.~A.}\ \bibnamefont {Liew}}, \bibinfo {author} {\bibfnamefont {S.}~\bibnamefont {Knappe}}, \bibinfo {author} {\bibfnamefont {J.}~\bibnamefont {Moreland}}, \bibinfo {author} {\bibfnamefont {H.}~\bibnamefont {Robinson}}, \bibinfo {author} {\bibfnamefont {L.}~\bibnamefont {Hollberg}},\ and\ \bibinfo {author} {\bibfnamefont {J.}~\bibnamefont {Kitching}},\ }\bibfield  {title} {\bibinfo {title} {Microfabricated alkali atom vapor cells},\ }\href {https://doi.org/10.1063/1.1691490} {\bibfield  {journal} {\bibinfo  {journal} {Applied Physics Letters}\ }\textbf {\bibinfo {volume} {84}},\ \bibinfo {pages} {2694} (\bibinfo {year} {2004})}\BibitemShut {NoStop}%
\bibitem [{\citenamefont {Maurice}\ \emph {et~al.}(2022)\citenamefont {Maurice}, \citenamefont {Carle}, \citenamefont {Keshavarzi}, \citenamefont {Chutani}, \citenamefont {Queste}, \citenamefont {Gauthier-Manuel}, \citenamefont {Cote}, \citenamefont {Vicarini}, \citenamefont {Abdel~Hafiz}, \citenamefont {Boudot},\ and\ \citenamefont {Passilly}}]{Maurice2022}%
  \BibitemOpen
  \bibfield  {author} {\bibinfo {author} {\bibfnamefont {V.}~\bibnamefont {Maurice}}, \bibinfo {author} {\bibfnamefont {C.}~\bibnamefont {Carle}}, \bibinfo {author} {\bibfnamefont {S.}~\bibnamefont {Keshavarzi}}, \bibinfo {author} {\bibfnamefont {R.}~\bibnamefont {Chutani}}, \bibinfo {author} {\bibfnamefont {S.}~\bibnamefont {Queste}}, \bibinfo {author} {\bibfnamefont {L.}~\bibnamefont {Gauthier-Manuel}}, \bibinfo {author} {\bibfnamefont {J.-M.}\ \bibnamefont {Cote}}, \bibinfo {author} {\bibfnamefont {R.}~\bibnamefont {Vicarini}}, \bibinfo {author} {\bibfnamefont {M.}~\bibnamefont {Abdel~Hafiz}}, \bibinfo {author} {\bibfnamefont {R.}~\bibnamefont {Boudot}},\ and\ \bibinfo {author} {\bibfnamefont {N.}~\bibnamefont {Passilly}},\ }\bibfield  {title} {\bibinfo {title} {Wafer-level vapor cells filled with laser-actuated hermetic seals for integrated atomic devices},\ }\href {https://doi.org/10.1038/s41378-022-00468-x} {\bibfield  {journal} {\bibinfo  {journal} {Microsystems \& Nanoengineering}\ }\textbf {\bibinfo
  {volume} {8}},\ \bibinfo {pages} {129} (\bibinfo {year} {2022})}\BibitemShut {NoStop}%
\bibitem [{\citenamefont {Cooper}\ \emph {et~al.}(2023{\natexlab{a}})\citenamefont {Cooper}, \citenamefont {Madkhaly}, \citenamefont {Johnson}, \citenamefont {Hopton}, \citenamefont {Baldolini},\ and\ \citenamefont {Hackerm\"{u}ller}}]{Cooper2023}%
  \BibitemOpen
  \bibfield  {author} {\bibinfo {author} {\bibfnamefont {N.}~\bibnamefont {Cooper}}, \bibinfo {author} {\bibfnamefont {S.}~\bibnamefont {Madkhaly}}, \bibinfo {author} {\bibfnamefont {D.}~\bibnamefont {Johnson}}, \bibinfo {author} {\bibfnamefont {B.}~\bibnamefont {Hopton}}, \bibinfo {author} {\bibfnamefont {D.}~\bibnamefont {Baldolini}},\ and\ \bibinfo {author} {\bibfnamefont {L.}~\bibnamefont {Hackerm\"{u}ller}},\ }\bibfield  {title} {\bibinfo {title} {Dual-frequency doppler-free spectroscopy for simultaneous laser stabilization in compact atomic physics experiments},\ }\href {https://doi.org/https://doi.org/10.1103/PhysRevA.108.013521} {\bibfield  {journal} {\bibinfo  {journal} {Phys. Rev. A}\ }\textbf {\bibinfo {volume} {108}},\ \bibinfo {pages} {013521} (\bibinfo {year} {2023}{\natexlab{a}})}\BibitemShut {NoStop}%
\bibitem [{\citenamefont {Martinez}\ \emph {et~al.}(2023)\citenamefont {Martinez}, \citenamefont {Li}, \citenamefont {Staron}, \citenamefont {Kitching}, \citenamefont {Raman},\ and\ \citenamefont {McGehee}}]{Martinez2023}%
  \BibitemOpen
  \bibfield  {author} {\bibinfo {author} {\bibfnamefont {G.~D.}\ \bibnamefont {Martinez}}, \bibinfo {author} {\bibfnamefont {C.}~\bibnamefont {Li}}, \bibinfo {author} {\bibfnamefont {A.}~\bibnamefont {Staron}}, \bibinfo {author} {\bibfnamefont {J.}~\bibnamefont {Kitching}}, \bibinfo {author} {\bibfnamefont {C.}~\bibnamefont {Raman}},\ and\ \bibinfo {author} {\bibfnamefont {W.~R.}\ \bibnamefont {McGehee}},\ }\bibfield  {title} {\bibinfo {title} {A chip-scale atomic beam clock},\ }\href {https://doi.org/https://doi.org/10.1038/s41467-023-39166-1} {\bibfield  {journal} {\bibinfo  {journal} {Nature Communications}\ }\textbf {\bibinfo {volume} {14}},\ \bibinfo {pages} {3501} (\bibinfo {year} {2023})}\BibitemShut {NoStop}%
\bibitem [{\citenamefont {Vilas}\ \emph {et~al.}(2022)\citenamefont {Vilas}, \citenamefont {Hallas}, \citenamefont {Anderegg}, \citenamefont {Robichaud}, \citenamefont {Winnicki}, \citenamefont {Mitra},\ and\ \citenamefont {Doyle}}]{Vilas_2022}%
  \BibitemOpen
  \bibfield  {author} {\bibinfo {author} {\bibfnamefont {N.~B.}\ \bibnamefont {Vilas}}, \bibinfo {author} {\bibfnamefont {C.}~\bibnamefont {Hallas}}, \bibinfo {author} {\bibfnamefont {L.}~\bibnamefont {Anderegg}}, \bibinfo {author} {\bibfnamefont {P.}~\bibnamefont {Robichaud}}, \bibinfo {author} {\bibfnamefont {A.}~\bibnamefont {Winnicki}}, \bibinfo {author} {\bibfnamefont {D.}~\bibnamefont {Mitra}},\ and\ \bibinfo {author} {\bibfnamefont {J.~M.}\ \bibnamefont {Doyle}},\ }\bibfield  {title} {\bibinfo {title} {Magneto-optical trapping and sub-doppler cooling of a polyatomic molecule},\ }\href {https://doi.org/10.1038/s41586-022-04620-5} {\bibfield  {journal} {\bibinfo  {journal} {Nature}\ }\textbf {\bibinfo {volume} {606}},\ \bibinfo {pages} {70} (\bibinfo {year} {2022})}\BibitemShut {NoStop}%
\bibitem [{\citenamefont {H\"{a}ffner}\ \emph {et~al.}(2008)\citenamefont {H\"{a}ffner}, \citenamefont {Roos},\ and\ \citenamefont {Blatt}}]{HAFFNER_2008}%
  \BibitemOpen
  \bibfield  {author} {\bibinfo {author} {\bibfnamefont {H.}~\bibnamefont {H\"{a}ffner}}, \bibinfo {author} {\bibfnamefont {C.}~\bibnamefont {Roos}},\ and\ \bibinfo {author} {\bibfnamefont {R.}~\bibnamefont {Blatt}},\ }\bibfield  {title} {\bibinfo {title} {Quantum computing with trapped ions},\ }\href {https://doi.org/10.1016/j.physrep.2008.09.003} {\bibfield  {journal} {\bibinfo  {journal} {Physics Reports}\ }\textbf {\bibinfo {volume} {469}},\ \bibinfo {pages} {155} (\bibinfo {year} {2008})}\BibitemShut {NoStop}%
\bibitem [{\citenamefont {Marciniak}\ \emph {et~al.}(2022)\citenamefont {Marciniak}, \citenamefont {Feldker}, \citenamefont {Pogorelov}, \citenamefont {Kaubrugger}, \citenamefont {Vasilyev}, \citenamefont {van Bijnen}, \citenamefont {Schindler}, \citenamefont {Zoller}, \citenamefont {Blatt},\ and\ \citenamefont {Monz}}]{Marciniak2023}%
  \BibitemOpen
  \bibfield  {author} {\bibinfo {author} {\bibfnamefont {C.~D.}\ \bibnamefont {Marciniak}}, \bibinfo {author} {\bibfnamefont {T.}~\bibnamefont {Feldker}}, \bibinfo {author} {\bibfnamefont {I.}~\bibnamefont {Pogorelov}}, \bibinfo {author} {\bibfnamefont {R.}~\bibnamefont {Kaubrugger}}, \bibinfo {author} {\bibfnamefont {D.~V.}\ \bibnamefont {Vasilyev}}, \bibinfo {author} {\bibfnamefont {R.}~\bibnamefont {van Bijnen}}, \bibinfo {author} {\bibfnamefont {P.}~\bibnamefont {Schindler}}, \bibinfo {author} {\bibfnamefont {P.}~\bibnamefont {Zoller}}, \bibinfo {author} {\bibfnamefont {R.}~\bibnamefont {Blatt}},\ and\ \bibinfo {author} {\bibfnamefont {T.}~\bibnamefont {Monz}},\ }\bibfield  {title} {\bibinfo {title} {Optimal metrology with programmable quantum sensors},\ }\href {https://doi.org/https://doi.org/10.1038/s41467-023-39166-1} {\bibfield  {journal} {\bibinfo  {journal} {Nature}\ }\textbf {\bibinfo {volume} {603}},\ \bibinfo {pages} {604} (\bibinfo {year} {2022})}\BibitemShut {NoStop}%
\bibitem [{\citenamefont {Fabricant}\ \emph {et~al.}(2023)\citenamefont {Fabricant}, \citenamefont {Novikova},\ and\ \citenamefont {Bison}}]{Fabricant_2023}%
  \BibitemOpen
  \bibfield  {author} {\bibinfo {author} {\bibfnamefont {A.}~\bibnamefont {Fabricant}}, \bibinfo {author} {\bibfnamefont {I.}~\bibnamefont {Novikova}},\ and\ \bibinfo {author} {\bibfnamefont {G.}~\bibnamefont {Bison}},\ }\bibfield  {title} {\bibinfo {title} {How to build a magnetometer with thermal atomic vapor: a tutorial},\ }\href {https://doi.org/10.1088/1367-2630/acb840} {\bibfield  {journal} {\bibinfo  {journal} {New Journal of Physics}\ }\textbf {\bibinfo {volume} {25}},\ \bibinfo {pages} {025001} (\bibinfo {year} {2023})}\BibitemShut {NoStop}%
\bibitem [{\citenamefont {Pross}\ \emph {et~al.}(2005)\citenamefont {Pross}, \citenamefont {Crisan}, \citenamefont {Bending}, \citenamefont {Mosser},\ and\ \citenamefont {Konczykowski}}]{pross2005second}%
  \BibitemOpen
  \bibfield  {author} {\bibinfo {author} {\bibfnamefont {A.}~\bibnamefont {Pross}}, \bibinfo {author} {\bibfnamefont {A.}~\bibnamefont {Crisan}}, \bibinfo {author} {\bibfnamefont {S.}~\bibnamefont {Bending}}, \bibinfo {author} {\bibfnamefont {V.}~\bibnamefont {Mosser}},\ and\ \bibinfo {author} {\bibfnamefont {M.}~\bibnamefont {Konczykowski}},\ }\bibfield  {title} {\bibinfo {title} {Second-generation quantum-well sensors for room-temperature scanning hall probe microscopy},\ }\href {https://doi.org/https://doi.org/10.1063/1.1887828} {\bibfield  {journal} {\bibinfo  {journal} {Journal of Applied Physics}\ }\textbf {\bibinfo {volume} {97}},\ \bibinfo {pages} {0964105} (\bibinfo {year} {2005})}\BibitemShut {NoStop}%
\bibitem [{\citenamefont {Behzadirad}\ \emph {et~al.}(2021)\citenamefont {Behzadirad}, \citenamefont {Mecholdt}, \citenamefont {Randall}, \citenamefont {Ballard}, \citenamefont {Owen}, \citenamefont {Rishinaramangalam}, \citenamefont {Reum}, \citenamefont {Gotszalk}, \citenamefont {Feezell}, \citenamefont {Rangelow} \emph {et~al.}}]{behzadirad2021advanced}%
  \BibitemOpen
  \bibfield  {author} {\bibinfo {author} {\bibfnamefont {M.}~\bibnamefont {Behzadirad}}, \bibinfo {author} {\bibfnamefont {S.}~\bibnamefont {Mecholdt}}, \bibinfo {author} {\bibfnamefont {J.~N.}\ \bibnamefont {Randall}}, \bibinfo {author} {\bibfnamefont {J.~B.}\ \bibnamefont {Ballard}}, \bibinfo {author} {\bibfnamefont {J.}~\bibnamefont {Owen}}, \bibinfo {author} {\bibfnamefont {A.~K.}\ \bibnamefont {Rishinaramangalam}}, \bibinfo {author} {\bibfnamefont {A.}~\bibnamefont {Reum}}, \bibinfo {author} {\bibfnamefont {T.}~\bibnamefont {Gotszalk}}, \bibinfo {author} {\bibfnamefont {D.~F.}\ \bibnamefont {Feezell}}, \bibinfo {author} {\bibfnamefont {I.~W.}\ \bibnamefont {Rangelow}}, \emph {et~al.},\ }\bibfield  {title} {\bibinfo {title} {Advanced scanning probe nanolithography using gan nanowires},\ }\href {https://doi.org/https://doi.org/10.1021/acs.nanolett.1c00127} {\bibfield  {journal} {\bibinfo  {journal} {Nano Letters}\ }\textbf {\bibinfo {volume} {21}},\ \bibinfo {pages} {5493} (\bibinfo {year}
  {2021})}\BibitemShut {NoStop}%
\bibitem [{\citenamefont {P{\'e}tremand}\ \emph {et~al.}(2012)\citenamefont {P{\'e}tremand}, \citenamefont {Affolderbach}, \citenamefont {Straessle}, \citenamefont {Pellaton}, \citenamefont {Briand}, \citenamefont {Mileti},\ and\ \citenamefont {de~Rooij}}]{petremand2012microfabricated}%
  \BibitemOpen
  \bibfield  {author} {\bibinfo {author} {\bibfnamefont {Y.}~\bibnamefont {P{\'e}tremand}}, \bibinfo {author} {\bibfnamefont {C.}~\bibnamefont {Affolderbach}}, \bibinfo {author} {\bibfnamefont {R.}~\bibnamefont {Straessle}}, \bibinfo {author} {\bibfnamefont {M.}~\bibnamefont {Pellaton}}, \bibinfo {author} {\bibfnamefont {D.}~\bibnamefont {Briand}}, \bibinfo {author} {\bibfnamefont {G.}~\bibnamefont {Mileti}},\ and\ \bibinfo {author} {\bibfnamefont {N.~F.}\ \bibnamefont {de~Rooij}},\ }\bibfield  {title} {\bibinfo {title} {Microfabricated rubidium vapour cell with a thick glass core for small-scale atomic clock applications},\ }\href {https://doi.org/10.1088/0960-1317/22/2/025013} {\bibfield  {journal} {\bibinfo  {journal} {Journal of Micromechanics and Microengineering}\ }\textbf {\bibinfo {volume} {22}},\ \bibinfo {pages} {025013} (\bibinfo {year} {2012})}\BibitemShut {NoStop}%
\bibitem [{\citenamefont {Shah}\ \emph {et~al.}(2007)\citenamefont {Shah}, \citenamefont {Knappe}, \citenamefont {Schwindt},\ and\ \citenamefont {Kitching}}]{shah2007subpicotesla}%
  \BibitemOpen
  \bibfield  {author} {\bibinfo {author} {\bibfnamefont {V.}~\bibnamefont {Shah}}, \bibinfo {author} {\bibfnamefont {S.}~\bibnamefont {Knappe}}, \bibinfo {author} {\bibfnamefont {P.~D.}\ \bibnamefont {Schwindt}},\ and\ \bibinfo {author} {\bibfnamefont {J.}~\bibnamefont {Kitching}},\ }\bibfield  {title} {\bibinfo {title} {Subpicotesla atomic magnetometry with a microfabricated vapour cell},\ }\href {https://doi.org/https://doi.org/10.1038/nphoton.2007.201} {\bibfield  {journal} {\bibinfo  {journal} {Nature Photonics}\ }\textbf {\bibinfo {volume} {1}},\ \bibinfo {pages} {649} (\bibinfo {year} {2007})}\BibitemShut {NoStop}%
\bibitem [{\citenamefont {Gong}\ \emph {et~al.}(2006)\citenamefont {Gong}, \citenamefont {Jau}, \citenamefont {Jensen},\ and\ \citenamefont {Happer}}]{Gong2006}%
  \BibitemOpen
  \bibfield  {author} {\bibinfo {author} {\bibfnamefont {F.}~\bibnamefont {Gong}}, \bibinfo {author} {\bibfnamefont {Y.-Y.}\ \bibnamefont {Jau}}, \bibinfo {author} {\bibfnamefont {K.}~\bibnamefont {Jensen}},\ and\ \bibinfo {author} {\bibfnamefont {W.}~\bibnamefont {Happer}},\ }\bibfield  {title} {\bibinfo {title} {Electrolytic fabrication of atomic clock cells},\ }\href {https://doi.org/https://doi.org/10.1063/1.2219730} {\bibfield  {journal} {\bibinfo  {journal} {Rev. Sci. Instrum.}\ }\textbf {\bibinfo {volume} {77}},\ \bibinfo {pages} {076101} (\bibinfo {year} {2006})}\BibitemShut {NoStop}%
\bibitem [{\citenamefont {Griffith}\ \emph {et~al.}(2010)\citenamefont {Griffith}, \citenamefont {Knappe},\ and\ \citenamefont {Kitching}}]{serf}%
  \BibitemOpen
  \bibfield  {author} {\bibinfo {author} {\bibfnamefont {W.~C.}\ \bibnamefont {Griffith}}, \bibinfo {author} {\bibfnamefont {S.}~\bibnamefont {Knappe}},\ and\ \bibinfo {author} {\bibfnamefont {J.}~\bibnamefont {Kitching}},\ }\bibfield  {title} {\bibinfo {title} {Femtotesla atomic magnetometry in a microfabricated vapor cell},\ }\href {https://doi.org/https://doi.org/10.1364/OE.18.027167} {\bibfield  {journal} {\bibinfo  {journal} {Optics Express}\ }\textbf {\bibinfo {volume} {18}},\ \bibinfo {pages} {27167} (\bibinfo {year} {2010})}\BibitemShut {NoStop}%
\bibitem [{\citenamefont {Lee}\ \emph {et~al.}(2004)\citenamefont {Lee}, \citenamefont {Guo}, \citenamefont {Radhakrishnam}, \citenamefont {Lal}, \citenamefont {Szekely}, \citenamefont {McClelland},\ and\ \citenamefont {Pisano}}]{Lee2004}%
  \BibitemOpen
  \bibfield  {author} {\bibinfo {author} {\bibfnamefont {C.~H.}\ \bibnamefont {Lee}}, \bibinfo {author} {\bibfnamefont {H.}~\bibnamefont {Guo}}, \bibinfo {author} {\bibfnamefont {S.}~\bibnamefont {Radhakrishnam}}, \bibinfo {author} {\bibfnamefont {A.}~\bibnamefont {Lal}}, \bibinfo {author} {\bibfnamefont {C.}~\bibnamefont {Szekely}}, \bibinfo {author} {\bibfnamefont {T.}~\bibnamefont {McClelland}},\ and\ \bibinfo {author} {\bibfnamefont {A.~P.}\ \bibnamefont {Pisano}},\ }in\ \href@noop {} {\emph {\bibinfo {booktitle} {Proc. Solid-State Sensors, Actuators and Microsystems Workshop (Hilton Head Island, SC, 6–10 June 2004)}}}\ (\bibinfo {year} {2004})\BibitemShut {NoStop}%
\bibitem [{\citenamefont {Luo}\ \emph {et~al.}(2017)\citenamefont {Luo}, \citenamefont {Gilbert}, \citenamefont {Qu}, \citenamefont {Landers}, \citenamefont {Bristow},\ and\ \citenamefont {Kinzel}}]{luo2017additive}%
  \BibitemOpen
  \bibfield  {author} {\bibinfo {author} {\bibfnamefont {J.}~\bibnamefont {Luo}}, \bibinfo {author} {\bibfnamefont {L.~J.}\ \bibnamefont {Gilbert}}, \bibinfo {author} {\bibfnamefont {C.}~\bibnamefont {Qu}}, \bibinfo {author} {\bibfnamefont {R.~G.}\ \bibnamefont {Landers}}, \bibinfo {author} {\bibfnamefont {D.~A.}\ \bibnamefont {Bristow}},\ and\ \bibinfo {author} {\bibfnamefont {E.~C.}\ \bibnamefont {Kinzel}},\ }\bibfield  {title} {\bibinfo {title} {Additive manufacturing of transparent soda-lime glass using a filament-fed process},\ }\href {https://doi.org/https://doi.org/10.1115/1.4035182} {\bibfield  {journal} {\bibinfo  {journal} {Journal of Manufacturing Science and Engineering}\ }\textbf {\bibinfo {volume} {139}},\ \bibinfo {pages} {061006} (\bibinfo {year} {2017})}\BibitemShut {NoStop}%
\bibitem [{\citenamefont {Inamura}\ \emph {et~al.}(2018)\citenamefont {Inamura}, \citenamefont {Stern}, \citenamefont {Lizardo}, \citenamefont {Houk},\ and\ \citenamefont {Oxman}}]{inamura2018additive}%
  \BibitemOpen
  \bibfield  {author} {\bibinfo {author} {\bibfnamefont {C.}~\bibnamefont {Inamura}}, \bibinfo {author} {\bibfnamefont {M.}~\bibnamefont {Stern}}, \bibinfo {author} {\bibfnamefont {D.}~\bibnamefont {Lizardo}}, \bibinfo {author} {\bibfnamefont {P.}~\bibnamefont {Houk}},\ and\ \bibinfo {author} {\bibfnamefont {N.}~\bibnamefont {Oxman}},\ }\bibfield  {title} {\bibinfo {title} {Additive manufacturing of transparent glass structures},\ }\href {https://doi.org/https://doi.org/10.1089/3dp.2018.0157} {\bibfield  {journal} {\bibinfo  {journal} {3D Printing and Additive Manufacturing}\ }\textbf {\bibinfo {volume} {5}},\ \bibinfo {pages} {269} (\bibinfo {year} {2018})}\BibitemShut {NoStop}%
\bibitem [{\citenamefont {Datsiou}\ \emph {et~al.}(2019)\citenamefont {Datsiou}, \citenamefont {Saleh}, \citenamefont {Spirrett}, \citenamefont {Goodridge}, \citenamefont {Ashcroft},\ and\ \citenamefont {Eustice}}]{datsiou2019additive}%
  \BibitemOpen
  \bibfield  {author} {\bibinfo {author} {\bibfnamefont {K.~C.}\ \bibnamefont {Datsiou}}, \bibinfo {author} {\bibfnamefont {E.}~\bibnamefont {Saleh}}, \bibinfo {author} {\bibfnamefont {F.}~\bibnamefont {Spirrett}}, \bibinfo {author} {\bibfnamefont {R.}~\bibnamefont {Goodridge}}, \bibinfo {author} {\bibfnamefont {I.}~\bibnamefont {Ashcroft}},\ and\ \bibinfo {author} {\bibfnamefont {D.}~\bibnamefont {Eustice}},\ }\bibfield  {title} {\bibinfo {title} {Additive manufacturing of glass with laser powder bed fusion},\ }\href {https://doi.org/https://doi.org/10.1111/jace.16440} {\bibfield  {journal} {\bibinfo  {journal} {Journal of the American Ceramic Society}\ }\textbf {\bibinfo {volume} {102}},\ \bibinfo {pages} {4410} (\bibinfo {year} {2019})}\BibitemShut {NoStop}%
\bibitem [{\citenamefont {Lei}\ \emph {et~al.}(2019)\citenamefont {Lei}, \citenamefont {Hong}, \citenamefont {Zhang}, \citenamefont {Peng},\ and\ \citenamefont {Xiao}}]{lei2019additive}%
  \BibitemOpen
  \bibfield  {author} {\bibinfo {author} {\bibfnamefont {J.}~\bibnamefont {Lei}}, \bibinfo {author} {\bibfnamefont {Y.}~\bibnamefont {Hong}}, \bibinfo {author} {\bibfnamefont {Q.}~\bibnamefont {Zhang}}, \bibinfo {author} {\bibfnamefont {F.}~\bibnamefont {Peng}},\ and\ \bibinfo {author} {\bibfnamefont {H.}~\bibnamefont {Xiao}},\ }\bibfield  {title} {\bibinfo {title} {Additive manufacturing of fused silica glass using direct laser melting},\ }in\ \href@noop {} {\emph {\bibinfo {booktitle} {CLEO: Applications and Technology}}}\ (\bibinfo {organization} {Optica Publishing Group},\ \bibinfo {year} {2019})\ pp.\ \bibinfo {pages} {AW3I--4}\BibitemShut {NoStop}%
\bibitem [{\citenamefont {Sasan}\ \emph {et~al.}(2020)\citenamefont {Sasan}, \citenamefont {Lange}, \citenamefont {Yee}, \citenamefont {Dudukovic}, \citenamefont {Nguyen}, \citenamefont {Johnson}, \citenamefont {Herrera}, \citenamefont {Yoo}, \citenamefont {Sawvel}, \citenamefont {Ellis} \emph {et~al.}}]{sasan2020additive}%
  \BibitemOpen
  \bibfield  {author} {\bibinfo {author} {\bibfnamefont {K.}~\bibnamefont {Sasan}}, \bibinfo {author} {\bibfnamefont {A.}~\bibnamefont {Lange}}, \bibinfo {author} {\bibfnamefont {T.~D.}\ \bibnamefont {Yee}}, \bibinfo {author} {\bibfnamefont {N.}~\bibnamefont {Dudukovic}}, \bibinfo {author} {\bibfnamefont {D.~T.}\ \bibnamefont {Nguyen}}, \bibinfo {author} {\bibfnamefont {M.~A.}\ \bibnamefont {Johnson}}, \bibinfo {author} {\bibfnamefont {O.~D.}\ \bibnamefont {Herrera}}, \bibinfo {author} {\bibfnamefont {J.~H.}\ \bibnamefont {Yoo}}, \bibinfo {author} {\bibfnamefont {A.~M.}\ \bibnamefont {Sawvel}}, \bibinfo {author} {\bibfnamefont {M.~E.}\ \bibnamefont {Ellis}}, \emph {et~al.},\ }\bibfield  {title} {\bibinfo {title} {Additive manufacturing of optical quality germania--silica glasses},\ }\href {https://doi.org/https://doi.org/10.1021/acsami.9b21136} {\bibfield  {journal} {\bibinfo  {journal} {ACS applied materials \& interfaces}\ }\textbf {\bibinfo {volume} {12}},\ \bibinfo {pages} {6736} (\bibinfo {year}
  {2020})}\BibitemShut {NoStop}%
\bibitem [{\citenamefont {Kotz}\ \emph {et~al.}(2017)\citenamefont {Kotz}, \citenamefont {Arnold}, \citenamefont {Bauer}, \citenamefont {Schild}, \citenamefont {Keller}, \citenamefont {Sachsenheimer}, \citenamefont {Nargang}, \citenamefont {Richter}, \citenamefont {Helmer},\ and\ \citenamefont {Rapp}}]{kotz2017three}%
  \BibitemOpen
  \bibfield  {author} {\bibinfo {author} {\bibfnamefont {F.}~\bibnamefont {Kotz}}, \bibinfo {author} {\bibfnamefont {K.}~\bibnamefont {Arnold}}, \bibinfo {author} {\bibfnamefont {W.}~\bibnamefont {Bauer}}, \bibinfo {author} {\bibfnamefont {D.}~\bibnamefont {Schild}}, \bibinfo {author} {\bibfnamefont {N.}~\bibnamefont {Keller}}, \bibinfo {author} {\bibfnamefont {K.}~\bibnamefont {Sachsenheimer}}, \bibinfo {author} {\bibfnamefont {T.~M.}\ \bibnamefont {Nargang}}, \bibinfo {author} {\bibfnamefont {C.}~\bibnamefont {Richter}}, \bibinfo {author} {\bibfnamefont {D.}~\bibnamefont {Helmer}},\ and\ \bibinfo {author} {\bibfnamefont {B.~E.}\ \bibnamefont {Rapp}},\ }\bibfield  {title} {\bibinfo {title} {Three-dimensional printing of transparent fused silica glass},\ }\href {https://doi.org/https://doi.org/10.1038/s41467-019-09497-z} {\bibfield  {journal} {\bibinfo  {journal} {Nature}\ }\textbf {\bibinfo {volume} {544}},\ \bibinfo {pages} {337} (\bibinfo {year} {2017})}\BibitemShut {NoStop}%
\bibitem [{\citenamefont {Cooperstein}\ \emph {et~al.}(2018)\citenamefont {Cooperstein}, \citenamefont {Shukrun}, \citenamefont {Press}, \citenamefont {Kamyshny},\ and\ \citenamefont {Magdassi}}]{cooperstein2018additive}%
  \BibitemOpen
  \bibfield  {author} {\bibinfo {author} {\bibfnamefont {I.}~\bibnamefont {Cooperstein}}, \bibinfo {author} {\bibfnamefont {E.}~\bibnamefont {Shukrun}}, \bibinfo {author} {\bibfnamefont {O.}~\bibnamefont {Press}}, \bibinfo {author} {\bibfnamefont {A.}~\bibnamefont {Kamyshny}},\ and\ \bibinfo {author} {\bibfnamefont {S.}~\bibnamefont {Magdassi}},\ }\bibfield  {title} {\bibinfo {title} {Additive manufacturing of transparent silica glass from solutions},\ }\href {https://doi.org/https://doi.org/10.1021/acsami.8b03766} {\bibfield  {journal} {\bibinfo  {journal} {ACS applied materials \& interfaces}\ }\textbf {\bibinfo {volume} {10}},\ \bibinfo {pages} {18879} (\bibinfo {year} {2018})}\BibitemShut {NoStop}%
\bibitem [{\citenamefont {Moore}\ \emph {et~al.}(2020)\citenamefont {Moore}, \citenamefont {Barbera}, \citenamefont {Masania},\ and\ \citenamefont {Studart}}]{moore2020three}%
  \BibitemOpen
  \bibfield  {author} {\bibinfo {author} {\bibfnamefont {D.~G.}\ \bibnamefont {Moore}}, \bibinfo {author} {\bibfnamefont {L.}~\bibnamefont {Barbera}}, \bibinfo {author} {\bibfnamefont {K.}~\bibnamefont {Masania}},\ and\ \bibinfo {author} {\bibfnamefont {A.~R.}\ \bibnamefont {Studart}},\ }\bibfield  {title} {\bibinfo {title} {Three-dimensional printing of multicomponent glasses using phase-separating resins},\ }\href {https://doi.org/https://doi.org/10.1038/s41563-019-0525-y} {\bibfield  {journal} {\bibinfo  {journal} {Nature materials}\ }\textbf {\bibinfo {volume} {19}},\ \bibinfo {pages} {212} (\bibinfo {year} {2020})}\BibitemShut {NoStop}%
\bibitem [{\citenamefont {Toombs}\ \emph {et~al.}(2022)\citenamefont {Toombs}, \citenamefont {Luitz}, \citenamefont {Cook}, \citenamefont {Jenne}, \citenamefont {Li}, \citenamefont {Rapp}, \citenamefont {Kotz-Helmer},\ and\ \citenamefont {Taylor}}]{toombs2022volumetric}%
  \BibitemOpen
  \bibfield  {author} {\bibinfo {author} {\bibfnamefont {J.~T.}\ \bibnamefont {Toombs}}, \bibinfo {author} {\bibfnamefont {M.}~\bibnamefont {Luitz}}, \bibinfo {author} {\bibfnamefont {C.~C.}\ \bibnamefont {Cook}}, \bibinfo {author} {\bibfnamefont {S.}~\bibnamefont {Jenne}}, \bibinfo {author} {\bibfnamefont {C.~C.}\ \bibnamefont {Li}}, \bibinfo {author} {\bibfnamefont {B.~E.}\ \bibnamefont {Rapp}}, \bibinfo {author} {\bibfnamefont {F.}~\bibnamefont {Kotz-Helmer}},\ and\ \bibinfo {author} {\bibfnamefont {H.~K.}\ \bibnamefont {Taylor}},\ }\bibfield  {title} {\bibinfo {title} {Volumetric additive manufacturing of silica glass with microscale computed axial lithography},\ }\href {https://doi.org/https://www.science.org/doi/10.1126/science.abm6459} {\bibfield  {journal} {\bibinfo  {journal} {Science}\ }\textbf {\bibinfo {volume} {376}},\ \bibinfo {pages} {308} (\bibinfo {year} {2022})}\BibitemShut {NoStop}%
\bibitem [{\citenamefont {Kotz}\ \emph {et~al.}(2021)\citenamefont {Kotz}, \citenamefont {Quick}, \citenamefont {Risch}, \citenamefont {Martin}, \citenamefont {Hoose}, \citenamefont {Thiel}, \citenamefont {Helmer},\ and\ \citenamefont {Rapp}}]{kotz2021two}%
  \BibitemOpen
  \bibfield  {author} {\bibinfo {author} {\bibfnamefont {F.}~\bibnamefont {Kotz}}, \bibinfo {author} {\bibfnamefont {A.~S.}\ \bibnamefont {Quick}}, \bibinfo {author} {\bibfnamefont {P.}~\bibnamefont {Risch}}, \bibinfo {author} {\bibfnamefont {T.}~\bibnamefont {Martin}}, \bibinfo {author} {\bibfnamefont {T.}~\bibnamefont {Hoose}}, \bibinfo {author} {\bibfnamefont {M.}~\bibnamefont {Thiel}}, \bibinfo {author} {\bibfnamefont {D.}~\bibnamefont {Helmer}},\ and\ \bibinfo {author} {\bibfnamefont {B.~E.}\ \bibnamefont {Rapp}},\ }\bibfield  {title} {\bibinfo {title} {Two-photon polymerization of nanocomposites for the fabrication of transparent fused silica glass microstructures},\ }\href {https://doi.org/https://doi.org/10.1002/adma.202006341} {\bibfield  {journal} {\bibinfo  {journal} {Advanced Materials}\ }\textbf {\bibinfo {volume} {33}},\ \bibinfo {pages} {2006341} (\bibinfo {year} {2021})}\BibitemShut {NoStop}%
\bibitem [{\citenamefont {Han}\ \emph {et~al.}(2019)\citenamefont {Han}, \citenamefont {Yang}, \citenamefont {Fang},\ and\ \citenamefont {Lee}}]{han2019rapid}%
  \BibitemOpen
  \bibfield  {author} {\bibinfo {author} {\bibfnamefont {D.}~\bibnamefont {Han}}, \bibinfo {author} {\bibfnamefont {C.}~\bibnamefont {Yang}}, \bibinfo {author} {\bibfnamefont {N.~X.}\ \bibnamefont {Fang}},\ and\ \bibinfo {author} {\bibfnamefont {H.}~\bibnamefont {Lee}},\ }\bibfield  {title} {\bibinfo {title} {Rapid multi-material 3d printing with projection micro-stereolithography using dynamic fluidic control},\ }\href {https://doi.org/https://doi.org/10.1016/j.addma.2019.03.031} {\bibfield  {journal} {\bibinfo  {journal} {Additive Manufacturing}\ }\textbf {\bibinfo {volume} {27}},\ \bibinfo {pages} {606} (\bibinfo {year} {2019})}\BibitemShut {NoStop}%
\bibitem [{\citenamefont {Cooper}\ \emph {et~al.}(2023{\natexlab{b}})\citenamefont {Cooper}, \citenamefont {Madkhaly}, \citenamefont {Johnson}, \citenamefont {Hopton}, \citenamefont {Baldolini},\ and\ \citenamefont {Hackerm\"uller}}]{CooperDualFreq}%
  \BibitemOpen
  \bibfield  {author} {\bibinfo {author} {\bibfnamefont {N.}~\bibnamefont {Cooper}}, \bibinfo {author} {\bibfnamefont {S.}~\bibnamefont {Madkhaly}}, \bibinfo {author} {\bibfnamefont {D.}~\bibnamefont {Johnson}}, \bibinfo {author} {\bibfnamefont {B.}~\bibnamefont {Hopton}}, \bibinfo {author} {\bibfnamefont {D.}~\bibnamefont {Baldolini}},\ and\ \bibinfo {author} {\bibfnamefont {L.}~\bibnamefont {Hackerm\"uller}},\ }\bibfield  {title} {\bibinfo {title} {Dual-frequency doppler-free spectroscopy for simultaneous laser stabilization in compact atomic physics experiments},\ }\href {https://doi.org/10.1103/PhysRevA.108.013521} {\bibfield  {journal} {\bibinfo  {journal} {Phys. Rev. A}\ }\textbf {\bibinfo {volume} {108}},\ \bibinfo {pages} {013521} (\bibinfo {year} {2023}{\natexlab{b}})}\BibitemShut {NoStop}%
\bibitem [{\citenamefont {Brookes}\ \emph {et~al.}(2022)\citenamefont {Brookes}, \citenamefont {Legget}, \citenamefont {Rea}, \citenamefont {Hill}, \citenamefont {Holmes}, \citenamefont {Boto},\ and\ \citenamefont {Bowtell}}]{Brookes2022}%
  \BibitemOpen
  \bibfield  {author} {\bibinfo {author} {\bibfnamefont {M.~J.}\ \bibnamefont {Brookes}}, \bibinfo {author} {\bibfnamefont {J.}~\bibnamefont {Legget}}, \bibinfo {author} {\bibfnamefont {M.}~\bibnamefont {Rea}}, \bibinfo {author} {\bibfnamefont {R.~M.}\ \bibnamefont {Hill}}, \bibinfo {author} {\bibfnamefont {N.}~\bibnamefont {Holmes}}, \bibinfo {author} {\bibfnamefont {E.}~\bibnamefont {Boto}},\ and\ \bibinfo {author} {\bibfnamefont {R.}~\bibnamefont {Bowtell}},\ }\bibfield  {title} {\bibinfo {title} {Magnetoencephalography with optically pumped magnetometers (opm-meg): the next generation of functional neuroimaging},\ }\href {https://doi.org/doi: 10.1016/j.tins.2022.05.008} {\bibfield  {journal} {\bibinfo  {journal} {Trends in Neurosciences}\ }\textbf {\bibinfo {volume} {45}},\ \bibinfo {pages} {621} (\bibinfo {year} {2022})}\BibitemShut {NoStop}%
\bibitem [{\citenamefont {Cai}\ \emph {et~al.}(2020)\citenamefont {Cai}, \citenamefont {Guo}, \citenamefont {Wang}, \citenamefont {Li}, \citenamefont {Li}, \citenamefont {Qiu}, \citenamefont {Zhang},\ and\ \citenamefont {Lue}}]{Cai2020}%
  \BibitemOpen
  \bibfield  {author} {\bibinfo {author} {\bibfnamefont {P.}~\bibnamefont {Cai}}, \bibinfo {author} {\bibfnamefont {L.}~\bibnamefont {Guo}}, \bibinfo {author} {\bibfnamefont {H.}~\bibnamefont {Wang}}, \bibinfo {author} {\bibfnamefont {J.}~\bibnamefont {Li}}, \bibinfo {author} {\bibfnamefont {J.}~\bibnamefont {Li}}, \bibinfo {author} {\bibfnamefont {Y.}~\bibnamefont {Qiu}}, \bibinfo {author} {\bibfnamefont {Q.}~\bibnamefont {Zhang}},\ and\ \bibinfo {author} {\bibfnamefont {Q.}~\bibnamefont {Lue}},\ }\bibfield  {title} {\bibinfo {title} {Effects of slurry mixing methods and solid loading on 3d printed silica glass parts based on dlp stereolithography},\ }\href {https://doi.org/https://doi.org/10.1016/j.ceramint.2020.03.260} {\bibfield  {journal} {\bibinfo  {journal} {Ceramics International}\ }\textbf {\bibinfo {volume} {46}},\ \bibinfo {pages} {16833} (\bibinfo {year} {2020})}\BibitemShut {NoStop}%
\bibitem [{\citenamefont {Lucivero}\ \emph {et~al.}(2022)\citenamefont {Lucivero}, \citenamefont {Zanoni}, \citenamefont {Corrielli}, \citenamefont {Osellame},\ and\ \citenamefont {Mitchell}}]{Mitchell_2022}%
  \BibitemOpen
  \bibfield  {author} {\bibinfo {author} {\bibfnamefont {V.~G.}\ \bibnamefont {Lucivero}}, \bibinfo {author} {\bibfnamefont {A.}~\bibnamefont {Zanoni}}, \bibinfo {author} {\bibfnamefont {G.}~\bibnamefont {Corrielli}}, \bibinfo {author} {\bibfnamefont {R.}~\bibnamefont {Osellame}},\ and\ \bibinfo {author} {\bibfnamefont {M.~W.}\ \bibnamefont {Mitchell}},\ }\bibfield  {title} {\bibinfo {title} {Laser-written vapor cells for chip-scale atomic sensing and spectroscopy},\ }\href {https://doi.org/10.1364/OE.469296} {\bibfield  {journal} {\bibinfo  {journal} {Opt. Express}\ }\textbf {\bibinfo {volume} {30}},\ \bibinfo {pages} {27149} (\bibinfo {year} {2022})}\BibitemShut {NoStop}%
\bibitem [{\citenamefont {Yong-Taeg}\ \emph {et~al.}(2002)\citenamefont {Yong-Taeg}, \citenamefont {Fujino},\ and\ \citenamefont {Morinag}}]{Yong2002}%
  \BibitemOpen
  \bibfield  {author} {\bibinfo {author} {\bibfnamefont {O.}~\bibnamefont {Yong-Taeg}}, \bibinfo {author} {\bibfnamefont {S.}~\bibnamefont {Fujino}},\ and\ \bibinfo {author} {\bibfnamefont {K.}~\bibnamefont {Morinag}},\ }\bibfield  {title} {\bibinfo {title} {Fabrication of transparent silica glass by powder sintering},\ }\href {https://doi.org/https://doi.org/10.1016/S1468-6996(02)00030-X} {\bibfield  {journal} {\bibinfo  {journal} {Science and Technology of Advanced Materials}\ }\textbf {\bibinfo {volume} {3}},\ \bibinfo {pages} {297} (\bibinfo {year} {2002})}\BibitemShut {NoStop}%
\bibitem [{\citenamefont {Milonji\'{c}}\ \emph {et~al.}(2007)\citenamefont {Milonji\'{c}}, \citenamefont {\v{C}erovi\'{c}}, \citenamefont {\v{C}oke\v{s}a},\ and\ \citenamefont {Zec}}]{Milonjic2007}%
  \BibitemOpen
  \bibfield  {author} {\bibinfo {author} {\bibfnamefont {S.~K.}\ \bibnamefont {Milonji\'{c}}}, \bibinfo {author} {\bibfnamefont {L.~S.}\ \bibnamefont {\v{C}erovi\'{c}}}, \bibinfo {author} {\bibfnamefont {D.~M.}\ \bibnamefont {\v{C}oke\v{s}a}},\ and\ \bibinfo {author} {\bibfnamefont {S.}~\bibnamefont {Zec}},\ }\bibfield  {title} {\bibinfo {title} {The influence of cationic impurities in silica on its crystallization and point of zero charge},\ }\href {https://doi.org/https://doi.org/10.1016/S1468-6996(02)00030-X} {\bibfield  {journal} {\bibinfo  {journal} {Journal of colloid and interface science}\ }\textbf {\bibinfo {volume} {309}},\ \bibinfo {pages} {155} (\bibinfo {year} {2007})}\BibitemShut {NoStop}%
\bibitem [{\citenamefont {Shun}\ \emph {et~al.}(2021)\citenamefont {Shun}, \citenamefont {Hirai}, \citenamefont {Tabata},\ and\ \citenamefont {Tsuchiya}}]{Dispenserfilling}%
  \BibitemOpen
  \bibfield  {author} {\bibinfo {author} {\bibfnamefont {K.}~\bibnamefont {Shun}}, \bibinfo {author} {\bibfnamefont {Y.}~\bibnamefont {Hirai}}, \bibinfo {author} {\bibfnamefont {O.}~\bibnamefont {Tabata}},\ and\ \bibinfo {author} {\bibfnamefont {T.}~\bibnamefont {Tsuchiya}},\ }\bibfield  {title} {\bibinfo {title} {Microfabricated cs vapor cells filled with an on-chip dispensing component},\ }\href {https://doi.org/10.35848/1347-4065/abe203} {\bibfield  {journal} {\bibinfo  {journal} {Japanese Journal of Applied Physics}\ }\textbf {\bibinfo {volume} {60}},\ \bibinfo {pages} {SCCL01} (\bibinfo {year} {2021})}\BibitemShut {NoStop}%
\bibitem [{\citenamefont {Alvarez}\ \emph {et~al.}(2022)\citenamefont {Alvarez}, \citenamefont {Gomez}, \citenamefont {Coop}, \citenamefont {Zamora-Zamora}, \citenamefont {Mazzinghi},\ and\ \citenamefont {Mitchell}}]{Palacios2022}%
  \BibitemOpen
  \bibfield  {author} {\bibinfo {author} {\bibfnamefont {S.~P.}\ \bibnamefont {Alvarez}}, \bibinfo {author} {\bibfnamefont {P.}~\bibnamefont {Gomez}}, \bibinfo {author} {\bibfnamefont {S.}~\bibnamefont {Coop}}, \bibinfo {author} {\bibfnamefont {R.}~\bibnamefont {Zamora-Zamora}}, \bibinfo {author} {\bibfnamefont {C.}~\bibnamefont {Mazzinghi}},\ and\ \bibinfo {author} {\bibfnamefont {M.~W.}\ \bibnamefont {Mitchell}},\ }\bibfield  {title} {\bibinfo {title} {Single-domain bose condensate magnetometer achieves energy resolution per bandwidth below $\hbar$},\ }\href {https://doi.org/https://doi.org/10.1073/pnas.2115339119} {\bibfield  {journal} {\bibinfo  {journal} {PNAS}\ }\textbf {\bibinfo {volume} {119}},\ \bibinfo {pages} {e2115339119} (\bibinfo {year} {2022})}\BibitemShut {NoStop}%
\bibitem [{\citenamefont {Liu}\ \emph {et~al.}(2018)\citenamefont {Liu}, \citenamefont {Qian}, \citenamefont {Ni},\ and\ \citenamefont {Qiu}}]{Liu2018}%
  \BibitemOpen
  \bibfield  {author} {\bibinfo {author} {\bibfnamefont {C.}~\bibnamefont {Liu}}, \bibinfo {author} {\bibfnamefont {B.}~\bibnamefont {Qian}}, \bibinfo {author} {\bibfnamefont {R.}~\bibnamefont {Ni}},\ and\ \bibinfo {author} {\bibfnamefont {J.}~\bibnamefont {Qiu}},\ }\bibfield  {title} {\bibinfo {title} {3d printing of multicolor luminescent glass},\ }\href {https://doi.org/https://doi.org/10.1039/C8RA06706F} {\bibfield  {journal} {\bibinfo  {journal} {RSC Advances}\ }\textbf {\bibinfo {volume} {8}},\ \bibinfo {pages} {31564} (\bibinfo {year} {2018})}\BibitemShut {NoStop}%
\bibitem [{\citenamefont {Hu}\ \emph {et~al.}(2017)\citenamefont {Hu}, \citenamefont {Sun}, \citenamefont {Parmenter}, \citenamefont {Fay}, \citenamefont {Smith}, \citenamefont {Rance}, \citenamefont {He}, \citenamefont {Zhang}, \citenamefont {Liu},\ and\ \citenamefont {Irvine}}]{Hu2017}%
  \BibitemOpen
  \bibfield  {author} {\bibinfo {author} {\bibfnamefont {Q.}~\bibnamefont {Hu}}, \bibinfo {author} {\bibfnamefont {X.-Z.}\ \bibnamefont {Sun}}, \bibinfo {author} {\bibfnamefont {C.~D.}\ \bibnamefont {Parmenter}}, \bibinfo {author} {\bibfnamefont {M.~W.}\ \bibnamefont {Fay}}, \bibinfo {author} {\bibfnamefont {E.~F.}\ \bibnamefont {Smith}}, \bibinfo {author} {\bibfnamefont {G.~A.}\ \bibnamefont {Rance}}, \bibinfo {author} {\bibfnamefont {Y.}~\bibnamefont {He}}, \bibinfo {author} {\bibfnamefont {F.}~\bibnamefont {Zhang}}, \bibinfo {author} {\bibfnamefont {Y.}~\bibnamefont {Liu}},\ and\ \bibinfo {author} {\bibfnamefont {D.}~\bibnamefont {Irvine}},\ }\href {https://doi.org/https://doi.org/10.1038/s41598-017-17391-1} {\bibfield  {journal} {\bibinfo  {journal} {Scientific reports}\ }\textbf {\bibinfo {volume} {7}},\ \bibinfo {pages} {17150} (\bibinfo {year} {2017})}\BibitemShut {NoStop}%
\bibitem [{\citenamefont {Bharadwaj}\ and\ \citenamefont {Mukherji}(2014)}]{bharadwaj2014gold}%
\BibitemOpen
\bibfield  {author} {\bibinfo {author} {\bibfnamefont {R.}~\bibnamefont {Bharadwaj}}\ and\ \bibinfo {author} {\bibfnamefont {S.}~\bibnamefont {Mukherji}},\ }\bibfield  {title} {\bibinfo {title} {Gold nanoparticle coated u-bend fibre optic probe for localized surface plasmon resonance based detection of explosive vapours},\ }\href {https://doi.org/https://doi.org/10.1016/j.snb.2013.11.026} {\bibfield  {journal} {\bibinfo  {journal} {Sensors and Actuators B: Chemical}\ }\textbf {\bibinfo {volume} {192}},\ \bibinfo {pages} {804} (\bibinfo {year} {2014})}\BibitemShut {NoStop}%
\bibitem [{\citenamefont {Trindade}\ \emph {et~al.}(2021)\citenamefont {Trindade}, \citenamefont {Wang}, \citenamefont {Im}, \citenamefont {He}, \citenamefont {Balogh}, \citenamefont {Scurr}, \citenamefont {Gilmore}, \citenamefont {Tiddia}, \citenamefont {Saleh}, \citenamefont {Pervan} \emph {et~al.}}]{trindade2021residual}%
  \BibitemOpen
  \bibfield  {author} {\bibinfo {author} {\bibfnamefont {G.~F.}\ \bibnamefont {Trindade}}, \bibinfo {author} {\bibfnamefont {F.}~\bibnamefont {Wang}}, \bibinfo {author} {\bibfnamefont {J.}~\bibnamefont {Im}}, \bibinfo {author} {\bibfnamefont {Y.}~\bibnamefont {He}}, \bibinfo {author} {\bibfnamefont {A.}~\bibnamefont {Balogh}}, \bibinfo {author} {\bibfnamefont {D.}~\bibnamefont {Scurr}}, \bibinfo {author} {\bibfnamefont {I.}~\bibnamefont {Gilmore}}, \bibinfo {author} {\bibfnamefont {M.}~\bibnamefont {Tiddia}}, \bibinfo {author} {\bibfnamefont {E.}~\bibnamefont {Saleh}}, \bibinfo {author} {\bibfnamefont {D.}~\bibnamefont {Pervan}}, \emph {et~al.},\ }\bibfield  {title} {\bibinfo {title} {Residual polymer stabiliser causes anisotropic electrical conductivity during inkjet printing of metal nanoparticles},\ }\href {https://doi.org/https://doi.org/10.1038/s43246-021-00151-0} {\bibfield  {journal} {\bibinfo  {journal} {Communications Materials}\ }\textbf {\bibinfo {volume} {2}},\ \bibinfo {pages} {47} (\bibinfo
  {year} {2021})}\BibitemShut {NoStop}%
\bibitem [{\citenamefont {Wang}\ \emph {et~al.}(2021)\citenamefont {Wang}, \citenamefont {Gosling}, \citenamefont {Trindade}, \citenamefont {Rance}, \citenamefont {Makarovsky}, \citenamefont {Cottam}, \citenamefont {Kudrynskyi}, \citenamefont {Balanov}, \citenamefont {Greenaway}, \citenamefont {Wildman} \emph {et~al.}}]{wang2021inter}%
  \BibitemOpen
  \bibfield  {author} {\bibinfo {author} {\bibfnamefont {F.}~\bibnamefont {Wang}}, \bibinfo {author} {\bibfnamefont {J.~H.}\ \bibnamefont {Gosling}}, \bibinfo {author} {\bibfnamefont {G.~F.}\ \bibnamefont {Trindade}}, \bibinfo {author} {\bibfnamefont {G.~A.}\ \bibnamefont {Rance}}, \bibinfo {author} {\bibfnamefont {O.}~\bibnamefont {Makarovsky}}, \bibinfo {author} {\bibfnamefont {N.~D.}\ \bibnamefont {Cottam}}, \bibinfo {author} {\bibfnamefont {Z.}~\bibnamefont {Kudrynskyi}}, \bibinfo {author} {\bibfnamefont {A.~G.}\ \bibnamefont {Balanov}}, \bibinfo {author} {\bibfnamefont {M.~T.}\ \bibnamefont {Greenaway}}, \bibinfo {author} {\bibfnamefont {R.~D.}\ \bibnamefont {Wildman}}, \emph {et~al.},\ }\bibfield  {title} {\bibinfo {title} {Inter-flake quantum transport of electrons and holes in inkjet-printed graphene devices},\ }\href {https://doi.org/https://doi.org/10.1002/adfm.202170032} {\bibfield  {journal} {\bibinfo  {journal} {Advanced Functional Materials}\ }\textbf {\bibinfo {volume} {31}},\ \bibinfo {pages}
  {2007478} (\bibinfo {year} {2021})}\BibitemShut {NoStop}%
\bibitem [{\citenamefont {Austin}\ \emph {et~al.}(2023)\citenamefont {Austin}, \citenamefont {Cottam}, \citenamefont {Zhang}, \citenamefont {Wang}, \citenamefont {Gosling}, \citenamefont {Nelson-Dummet}, \citenamefont {James}, \citenamefont {Beton}, \citenamefont {Trindade}, \citenamefont {Zhou} \emph {et~al.}}]{austin2023photosensitisation}%
  \BibitemOpen
  \bibfield  {author} {\bibinfo {author} {\bibfnamefont {J.~S.}\ \bibnamefont {Austin}}, \bibinfo {author} {\bibfnamefont {N.~D.}\ \bibnamefont {Cottam}}, \bibinfo {author} {\bibfnamefont {C.}~\bibnamefont {Zhang}}, \bibinfo {author} {\bibfnamefont {F.}~\bibnamefont {Wang}}, \bibinfo {author} {\bibfnamefont {J.~H.}\ \bibnamefont {Gosling}}, \bibinfo {author} {\bibfnamefont {O.}~\bibnamefont {Nelson-Dummet}}, \bibinfo {author} {\bibfnamefont {T.~S.}\ \bibnamefont {James}}, \bibinfo {author} {\bibfnamefont {P.~H.}\ \bibnamefont {Beton}}, \bibinfo {author} {\bibfnamefont {G.~F.}\ \bibnamefont {Trindade}}, \bibinfo {author} {\bibfnamefont {Y.}~\bibnamefont {Zhou}}, \emph {et~al.},\ }\bibfield  {title} {\bibinfo {title} {Photosensitisation of inkjet printed graphene with stable all-inorganic perovskite nanocrystals},\ }\href {https://doi.org/https://doi.org/10.1039/D2NR06429D} {\bibfield  {journal} {\bibinfo  {journal} {Nanoscale}\ }\textbf {\bibinfo {volume} {15}},\ \bibinfo {pages} {2134} (\bibinfo {year}
  {2023})}\BibitemShut {NoStop}%
\bibitem [{\citenamefont {Faria-E-Silva}\ and\ \citenamefont {Pfeifer}(2017)}]{Faria2017}%
  \BibitemOpen
  \bibfield  {author} {\bibinfo {author} {\bibfnamefont {A.~L.}\ \bibnamefont {Faria-E-Silva}}\ and\ \bibinfo {author} {\bibfnamefont {C.~S.}\ \bibnamefont {Pfeifer}},\ }\bibfield  {title} {\bibinfo {title} {Impact of thio-urethane additive and filler type on light-transmission and depth of polymerization of dental composites},\ }\href {https://doi.org/https://doi.org/10.1016/j.dental.2017.07.020} {\bibfield  {journal} {\bibinfo  {journal} {Dental Materials}\ }\textbf {\bibinfo {volume} {33}},\ \bibinfo {pages} {1274} (\bibinfo {year} {2017})}\BibitemShut {NoStop}%
\bibitem [{\citenamefont {Zhao}\ \emph {et~al.}(2021)\citenamefont {Zhao}, \citenamefont {He}, \citenamefont {Trindade}, \citenamefont {Baumers}, \citenamefont {Irvine}, \citenamefont {Hague}, \citenamefont {Ashcroft},\ and\ \citenamefont {D.}}]{Zhao2021}%
  \BibitemOpen
  \bibfield  {author} {\bibinfo {author} {\bibfnamefont {P.}~\bibnamefont {Zhao}}, \bibinfo {author} {\bibfnamefont {Y.}~\bibnamefont {He}}, \bibinfo {author} {\bibfnamefont {G.~F.}\ \bibnamefont {Trindade}}, \bibinfo {author} {\bibfnamefont {M.}~\bibnamefont {Baumers}}, \bibinfo {author} {\bibfnamefont {D.~J.}\ \bibnamefont {Irvine}}, \bibinfo {author} {\bibfnamefont {R.~J.}\ \bibnamefont {Hague}}, \bibinfo {author} {\bibfnamefont {I.~A.}\ \bibnamefont {Ashcroft}},\ and\ \bibinfo {author} {\bibfnamefont {W.~R.}\ \bibnamefont {D.}},\ }\bibfield  {title} {\bibinfo {title} {Modelling the influence of uv curing strategies for optimisation of inkjet based 3d printing},\ }\href {https://doi.org/https://doi.org/10.1016/j.matdes.2021.109889} {\bibfield  {journal} {\bibinfo  {journal} {Materials \& Design}\ }\textbf {\bibinfo {volume} {208}},\ \bibinfo {pages} {109889} (\bibinfo {year} {2021})}\BibitemShut {NoStop}%
\end{thebibliography}
